 \documentclass[iicol,pdflatex,sn-basic]{sn-jnl}

\usepackage{graphicx}%
\usepackage{dcolumn}
\usepackage{multirow}%
\usepackage{amsmath,amssymb,amsfonts}%
\usepackage{amsthm}%
\usepackage{mathrsfs}%
\usepackage[title]{appendix}%
\usepackage{xcolor}%
\usepackage{textcomp}%
\usepackage{manyfoot}%
\usepackage{booktabs}%
\usepackage{algorithm}%
\usepackage{algorithmicx}%
\usepackage{algpseudocode}%
\usepackage{listings}%
\usepackage{mathrsfs} 
\usepackage{ulem}

\theoremstyle{thmstyleone}%
\theoremstyle{thmstyletwo}%

\theoremstyle{thmstylethree}%
\begin{document}

\title[]{Nonlinear synthetic Schlieren methods for free-surface topography measurement using telecentric imaging}

\author*[1,2,3]{\fnm{Shimin} \sur{Zhang}}\email{zsm031@sjtu.edu.cn}

\author[3]{\fnm{Frédéric} \sur{Moisy}}

\author[3]{\fnm{Wietze} \sur{Herreman}}

\author[1,2]{\fnm{Zhiliang} \sur{Lin}}

\affil[1]{State Key Laboratory of Ocean Engineering, Shanghai Jiao Tong University, Shanghai 200240, PR China}

\affil[2]{Marine Numerical Experiment Center, School of Ocean and Civil Engineering, Shanghai Jiao Tong University, Shanghai 200240, PR China}

\affil[3]{Université Paris-Saclay, CNRS, Laboratoire FAST, 91405 Orsay, France}


\abstract{Free-surface synthetic Schlieren (FS-SS) is a high-resolution, refraction-based optical technique for measuring the instantaneous elevation of a liquid interface. Under the assumptions of small amplitude, small slope, and small paraxial angle, the method yields a linear relationship between the gradient of the surface elevation and the apparent displacement field of a refracted pattern imaged through the surface. Here, we propose three new, nonlinear extensions of the FS-SS method that are specifically dedicated to telecentric imaging. Paraxial distortions are eliminated with a telecentric lens, thereby simplifying the optical model. This allows us to derive nonlinear surface reconstruction models that reach beyond the usual limits of small slope and small wave-magnitudes. We implement these nonlinear surface reconstruction algorithms and compare them to the original, linear reconstruction algorithm in three different experiments, using a solid glass lens, spreading oil drops and nonlinear Faraday waves. At the price of a few iterations, we can realise nonlinear surface reconstructions that are more precise, in particular when we reach high slopes or high amplitude regimes. We share a library that encodes these nonlinear surface reconstruction algorithms.}

\maketitle

\section{Introduction\label{sec:intro}}
The accurate measurement of free-surface topography is fundamental to fluid dynamics research, as it underpins the quantitative analysis of various interfacial phenomena. Over recent decades, several non-intrusive optical methods based on high-speed imaging have been developed for full three-dimensional (3-D) surface reconstruction \citep{falcon2022experiments,gomit2022free}, including Fourier transform profilometry \citep{cobelli2009global}, diffusing light profilometry or photography \citep{berhanu2013space,haudin2016experimental}, and synthetic Schlieren (also known as background-oriented Schlieren) \citep{kurata1990water,moisy2009synthetic,raffel2015background,vinnichenko2025background}. Among these, the free-surface synthetic Schlieren (FS-SS) method, proposed by \citet{moisy2009synthetic}, has emerged as a preeminent tool for instantaneous, wide-field measurements \citep{eddi2009unpredictable,moisy2012cross,abella2019detection,zhang2023wind,zhang2023pattern}. Its widespread adoption is attributed to a combination of experimental simplicity, high spatial resolution (approaching $1\,\mu\mathrm{m}$), and low computational cost.

The standard FS-SS method, as established by \citet{moisy2009synthetic}, relies on digital image correlation (DIC) to measure the apparent displacement field between a reference pattern and its image refracted through a deformed interface. Under a weak-deformation assumption, the measured displacement field is linearly related to the surface gradient, which is numerically inverted in the least-squares sense using finite-difference approximations to reconstruct the surface elevation. 

Despite its versatility, the standard FS-SS method has several limitations. (i) Under high surface curvature or with large surface-pattern distance, caustics may arise, preventing a one-to-one mapping between the reference and the refracted pattern. (ii) Being based on gradient integration, the method cannot detect uniform height variations or give absolute vertical positions. (iii)  The method is sensitive to mechanical vibrations, which introduce spurious noise and scaling artifacts in the presence of paraxial corrections. (iv) The linear correlation between the surface gradient and the displacement field relies on the assumptions of small paraxial angles, weak surface slopes, and small wave amplitudes relative to the surface-pattern distance. 

Considerable research over the past decade has sought to relax these limitations. The caustic problem is a fundamental limit for any optical method based on ray theory and there is no simple solution.  To address the absolute height-ambiguity problem, \citet{gomit2013free} introduced a multi-camera stereoscopic approach that increases hardware complexity and calibration demands. To mitigate environmental noise, \citet{damiano2016surface} and \citet{zhang2023pattern} developed multi-step processing protocols to filter vibration-induced artifacts and paraxial scaling distortions. To eliminate paraxial effects, \citet{metzmacher2022double} and \citet{mathis2026effet} used a telecentric lense, which introduces a limitation on the size of the region of interest. \citet{metzmacher2022double} also employed Fourier demodulation \citep{wildeman2018real} to achieve pixel-level resolution and proposed a double-pattern configuration enabling reconstruction in regimes of large surface slopes and unknown liquid depths. 
More recently, the FS-SS method has been extended to accommodate more complex experimental configurations. For instance, \citet{mermelstein2025modified} modified the technique to utilize arbitrarily sloped patterns, optimizing resolution across regions of varying wave amplitude. Additionally, \citet{zhang2025free} modeled light refraction through two deformed interfaces, creating a bed-independent formulation for measuring liquid surfaces over smoothly varying solid substrates. 

The linear relationship between the surface gradient and the optical displacement is a practical approximation that holds for small surface slopes and small wave magnitudes. When the wave height becomes comparable to the reference height, for example in thin film flows, this linear approximation is no longer sufficiently accurate. Another problem occurs when we want to measure the smaller, nonlinear wave content in weakly nonlinear waves. Optical non-linearities can be of the same order as hydrodynamical non-linearities and a linear FS-SS approach is then not sufficient to measure the smaller wave harmonics. These two examples suggest that it is of interest to develop new algorithms that go beyond the linearised approach.

A first weakly nonlinear FS-SS method was proposed by \citet{li2021single}. In their method, that includes paraxial distortions, the small amplitude constraint is relaxed but the method still assumes small slopes. The absolute local surface height is reconstructed via an iterative, Newton–Raphson solver.  This method was successfully applied to reconstruct surface topography in configurations where the mean liquid depth is not known a priori, such as dam-break flows.

In the present work, we also consider nonlinear extensions of the original linear FS–SS method, but unlike  \citet{li2021single}, we do not include paraxial effects. We focus on a synthetic schlieren measurements with  telecentric imaging, as considered by \citet{metzmacher2022double} and \citet{mathis2026effet}. By eliminating paraxial distorsions, telecentric imaging significantly simplifies the nonlinear ray-refraction model. This simplification allows us to propose three new iterative reconstruction algorithms with increasing levels of accuracy: 

\begin{itemize}

\item WNL-$h$ ($h$-based Weakly Non-Linear): This algorithm computes the small deviation $h$ of the interface with respect to a supposedly, much larger reference depth. It relies on the assumptions of (i) weak surface slope and (ii) small wave amplitude with respect to reference depth. In three iterations, it achieves third order accuracy in wave magnitude.  This method improves the first order, linear reconstruction, but the weakly nonlinear, higher order corrections are by definition small.  \\

\item WNL-$H$ ($H$-based Weakly Non-Linear): This algorithm is an iterative, weakly nonlinear method that computes the total surface height $H$. It still relies on the assumption of weak surface slope, but there is no restriction on wave amplitude with respect to reference depth. In one iteration it achieves second order accuracy, while fourth order accuracy can be reached in less than 5 iterations. \\

\item FNL-$H$ ($H$-based Fully Non-Linear): In this iterative algorithm, both assumptions of  small surface slope and small wave amplitude are relaxed. The iterative scheme is initialized using the second order, weakly nonlinear guess of the previous method, and convergence is achieved in as few iterations. 
\end{itemize}



All three algorithms are Jacobian-free and converge rapidly at low computational cost. We share them in an online library. The principal limitation of these nonlinear reconstruction methods is that they require knowledge of the absolute height at at least one point within the field of view, since telecentric lenses are insensitive to uniform surface elevations. This is an important difference with respect to \citet{li2021single} that could measure the absolute height without any supplementary constraints.  

  
The paper is organized as follows. In section \ref{sec:method}, we consider the refraction problem under orthographic projection and derive the different algorithms.  In section \ref{sec:practical}, we discuss the practical implementation. Section \ref{sec:numTest} shows a well-controlled numerical test. In section \ref{sec:expTest}, we apply the new reconstruction methods to three experimental configurations: a solid spherical lens, a spreading viscous drop and Faraday waves. Section \ref{sec:conclusion} summarizes the key findings and outlines the main contributions of the work.

\section{Refraction model formulation\label{sec:method}}

\subsection{Exact optical displacement \label{subsec:exact_model}}

\begin{figure}[t!]
    \centering
    \includegraphics[scale=1.1]{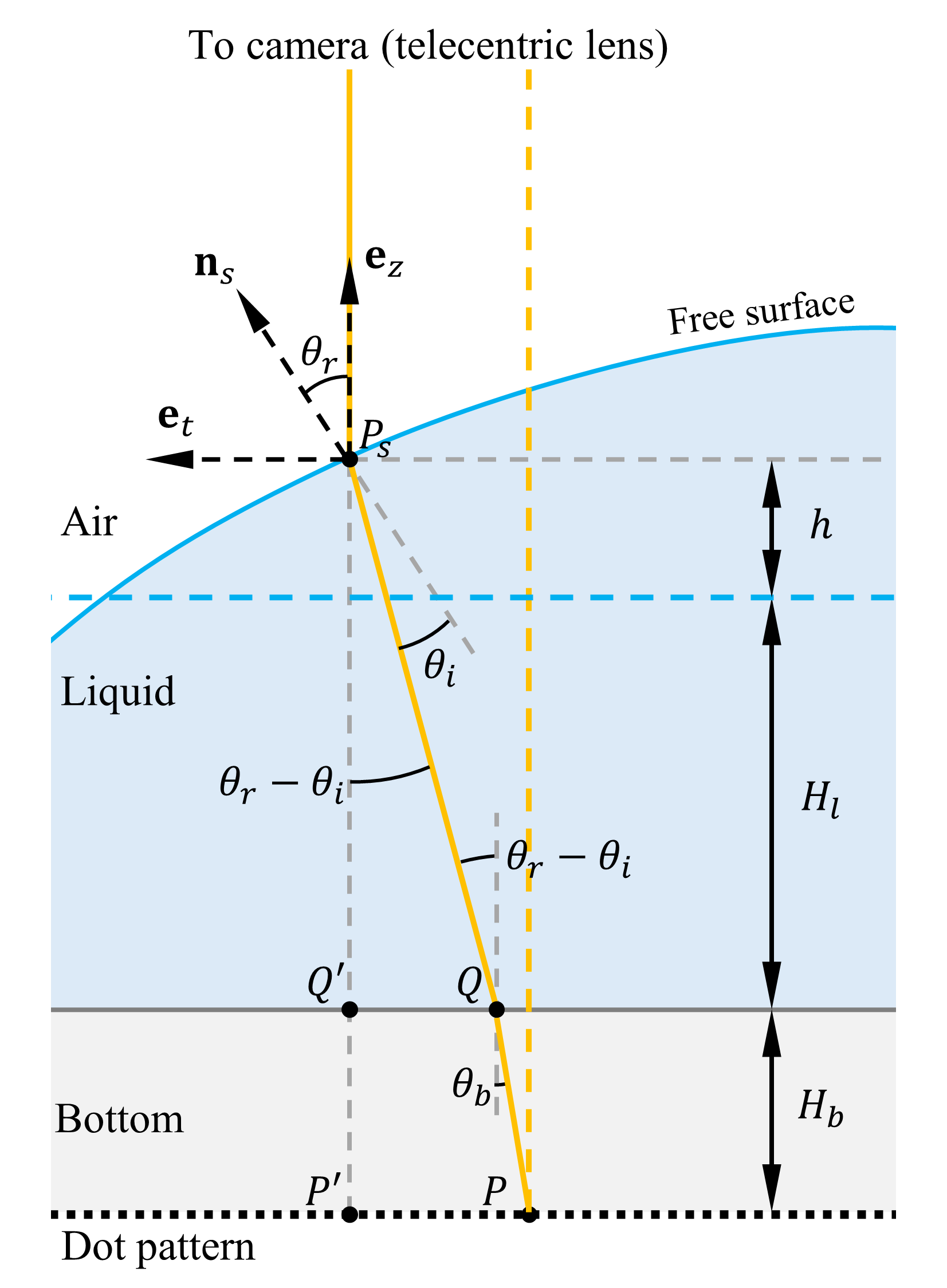}
  \caption{Schematic of light-ray refraction through a bottom solid spacer and a deformed air-liquid interface under telecentric observation}
    \label{fig:light_refraction}
\end{figure}

The optical configuration, illustrated in Fig.~\ref{fig:light_refraction}, comprises a camera equipped with a telecentric lens positioned above a fluid container with a reference dot pattern located bellow. We denote $H_l$ the mean liquid depth, $H_b$ the thickness of the bottom spacer (usually made of glass or perspex), and $h(x,y)$ the local surface height. The refractive indices of air, liquid, and the bottom material are denoted $n_a$, $n_l$, and $n_b$, respectively.

The orthographic projection inherent to the telecentric lens ensures that only rays emerging vertically from the interface are collected.  In the quiescent state (yellow dashed line in Fig.~\ref{fig:light_refraction}), a vertical ray originating from the reference dot $P$ is collected by the telecentric lens. Upon surface deformation $h$  (yellow solid line), the light ray originating from the same point $P$  crosses the solid-liquid and liquid-air interfaces at points $Q$ and $P_s$, respectively, and reaches the camera along the vertical direction $\mathbf{e}_z$. We denote $P'$ the apparent position of point $P$, defined as the intersection between the vertical line passing through $P_s$ and the horizontal plane of the dot pattern.

We note  $\delta\mathbf{r} = \mathbf{PP'}$ the apparent horizontal displacement, which is colinear with the unit vector $\mathbf{e}_t = -\boldsymbol{\nabla} h / \|\boldsymbol{\nabla} h\|$, where $\boldsymbol{\nabla} h$ is the surface gradient at point $P_s$. This displacement can be decomposed as $\delta\mathbf{r} = \delta\mathbf{r}_b + \delta\mathbf{r}_l$, where $\delta\mathbf{r}_l = \mathbf{QQ'}$ is the liquid-induced contribution and $\delta\mathbf{r}_b$ the bottom-spacer contribution (horizontal projection of $\mathbf{PQ}$). The relationship between these horizontal displacements and the surface geometry is governed by Snell's law, that is expressed using the layer thicknesses $H_b$ and $H_l$, the local elevation $h$, and the refraction angles $\theta_b$, $\theta_i$, and $\theta_r$ at each surface crossing.

From the geometric relationships in Fig.~\ref{fig:light_refraction}, the liquid-induced displacement is expressed as:
\begin{equation}
    \begin{split}
        \delta\mathbf{r}_l =& -\tan{(\theta_r - \theta_i)} (H_l+h) \mathbf{e}_t\\
        &= - \frac{\sin{(\theta_r - \theta_i)}}{\cos{(\theta_r - \theta_i)}} \frac{(H_l+h)\boldsymbol{\nabla} h}{\|\boldsymbol{\nabla} h\|},
    \end{split}
    \label{eq:defor}
\end{equation}
where $\theta_i$ and $\theta_r$ denote the incidence and refraction angles at the free surface, respectively. Given the verticality of the refracted ray and with the surface normal  $\mathbf{n}_s  =  (\mathbf{e}_z - \boldsymbol{\nabla} h) / \sqrt{1+\|\boldsymbol{\nabla} h\|^2}$, we can express the cosine and sine of the refraction angle $\theta_r$ as :
\begin{equation}
    \cos{\theta_r} =  \frac{1}{\sqrt{1+\|\boldsymbol{\nabla} h\|^2}}, \, \sin{\theta_r} = \frac{\|\boldsymbol{\nabla} h\|}{\sqrt{1+\|\boldsymbol{\nabla} h\|^2}}.
    \label{eq:theta_r}
\end{equation}
The angle of incidence $\theta_i$ is determined via Snell’s law, $\sin{\theta_i} = \eta \sin{\theta_r}$, where $\eta = n_a / n_l$ is the relative refractive index of air ($n_a$) to the liquid ($n_l$). This yields:
\begin{equation}
    \sin{\theta_i} = \frac{\eta\|\boldsymbol{\nabla} h\|}{\sqrt{1+\|\boldsymbol{\nabla} h\|^2}}, \, \cos{\theta_i} = \sqrt{\frac{1+(1-\eta^2)\|\boldsymbol{\nabla} h\|^2}{1+\|\boldsymbol{\nabla} h\|^2}},
    \label{eq:theta_i}
\end{equation}
and 
\begin{equation}
 \begin{split}
  \sin{(\theta_r-\theta_i)}=& \frac{(\sqrt{1+(1-\eta^2)\|\boldsymbol{\nabla} h\|^2}-\eta) \|\boldsymbol{\nabla} h\|}{1+\|\boldsymbol{\nabla} h\|^2} \\
                \cos{(\theta_r-\theta_i)}=& \frac{\sqrt{1+(1-\eta^2)\|\boldsymbol{\nabla} h\|^2}+\eta\|\boldsymbol{\nabla} h\|^2}{1+\|\boldsymbol{\nabla} h\|^2}.
\end{split}
 \label{eq:theta_rmi}
\end{equation}
Substituting these into Eq.~(\ref{eq:defor}) leads to the exact expression for the liquid-induced displacement:
\begin{equation}
    \delta\mathbf{r}_l = -\frac{\sqrt{1+(1-\eta^2)\|\boldsymbol{\nabla} h\|^2}-\eta}{\sqrt{1+(1-\eta^2)\|\boldsymbol{\nabla} h\|^2}+\eta\|\boldsymbol{\nabla} h\|^2} (H_l+h)\boldsymbol{\nabla} h.
    \label{eq:dr_l}
\end{equation}

Analogously, for the refraction at the bottom-liquid interface, Snell's law gives the incident angle $\theta_b$ through the relationship $\sin{\theta_b} = \eta_b \sin{(\theta_r - \theta_i)}$, where $\eta_b = n_l / n_b$ and $n_b$ is the refractive index of the bottom spacer. The bottom-induced displacement field is thus:
\begin{equation}
    \begin{split}
        \delta\mathbf{r}_b =& -H_b \tan{\theta_b} \mathbf{e}_t \\
        &= -\frac{\eta_b H_b \sin{(\theta_r - \theta_i)}}{\sqrt{1-\eta_b^2 \sin^2{(\theta_r - \theta_i)}}} \frac{\boldsymbol{\nabla} h}{\|\boldsymbol{\nabla} h\|}
    \end{split}
    \label{eq:dr_b}
\end{equation}
with $\sin(\theta_r - \theta_i)$  as specified in \eqref{eq:theta_rmi}. By adding $\delta\mathbf{r} = \delta\mathbf{r}_l + \delta\mathbf{r}_b$, the displacement field is uniquely related to the instantaneous liquid height $h$ and its gradient $\boldsymbol{\nabla} h$.

\subsection{Weak-slope approximation}
As \citet{li2021single}, we can simplify Eq.~(\ref{eq:dr_l}) by assuming a small surface slope $\|\boldsymbol{\nabla} h\| = \mathcal{O}(\varepsilon)$ and $\|\boldsymbol{\nabla} h\|^2 = \mathcal{O}(\varepsilon^2)$, where $\varepsilon$ is a small parameter. Using the Taylor series expansions $\sqrt{1 + \varepsilon^2} = 1 + \varepsilon^2/2 + \mathcal{O}(\varepsilon^4)$ and  
${(1 + \varepsilon^2)}^{-1} = 1 - \varepsilon^2 + \mathcal{O}(\varepsilon^4)$ and neglecting higher-order terms of $\mathcal{O}(\varepsilon^4)$, we simplify 
Eq.~(\ref{eq:dr_l}) for the liquid-induced displacement into
\begin{equation}
   \delta\mathbf{r}_l\approx-\alpha \left(1-\frac{1}{2}\eta \alpha\|\boldsymbol{\nabla} h\|^2\right)\left(H_l+h\right)\boldsymbol{\nabla} h,
\label{eq:gradcoe_simpleCase}
\end{equation}
where we denote $\alpha=1-\eta$. In a similar way, we simplify Eq.~(\ref{eq:dr_b}) for the bottom-induced displacement into:
\begin{equation}
        \delta \mathbf{r}_b\approx-\alpha\left(1-\frac{1}{2} \alpha \|\boldsymbol{\nabla} h\|^2 +\frac{1}{2} \alpha^2 \eta_b^2\|\boldsymbol{\nabla} h\|^2\right)\eta_b H_b \boldsymbol{\nabla} h.
\end{equation}
The total displacement $\delta \mathbf{r} = \delta\mathbf{r}_l + \delta\mathbf{r}_b$ is:
\begin{equation}
    \begin{split}
        \delta \mathbf{r}\approx&-\alpha \left(1-\frac{1}{2}\eta\alpha \|\boldsymbol{\nabla} h\|^2\right)\left(H_{\mathrm{eff}} +h\right)\boldsymbol{\nabla} h\\
        &+\frac{1}{2}\alpha^3 (1-\eta_b^2)\eta_b H_b \|\boldsymbol{\nabla} h\|^2 \boldsymbol{\nabla} h.
    \end{split}
    \label{eq:dr_h}
\end{equation}
We introduce $H_{\mathrm{eff}} = H_l + \eta_b H_b$ to denote the effective surface-pattern distance. The approximation used by \citet{li2021single} was of second order in slope ($ \mathcal{O}(\varepsilon^2)$) and would correspond here (without paraxial corrections) to taking $ \delta \mathbf{r} \approx  -\alpha \left(H_{\mathrm{eff}} +h\right)\boldsymbol{\nabla} h$. 

\subsection{Iterative surface reconstruction\label{subsec:reconstruction}}

\subsubsection{WNL-$h$ algorithm \label{subsubsec:WNLh}}

Our first nonlinear reconstruction algorithm relies, just as the original FS-SS method, on the assumptions that (i) the surface deformations have small slope, $ \| \boldsymbol{\nabla} h \| \ll 1$, and (ii) small amplitude with respect to the effective height, $|h|/H_{\mathrm{eff}} \ll 1$.  In this double limit, we can propose a perturbative, weakly nonlinear expansion of the surface elevation:
\begin{equation}
    h = h^{(1)} + h^{(2)} + h^{(3)} + \mathcal{O}(\varepsilon^4),
\end{equation}
where each term $h^{(i)}$ scales as $\mathcal{O}(\varepsilon^i)$ relative to the effective depth $H_{\mathrm{eff}}$. By substituting this expansion into Eq.~\eqref{eq:dr_h} and performing order-by-order matching, we derive a hierarchical set of linear equations that can be solved, order by order. This framework constitutes what we call, the weakly nonlinear $h$-based (WNL-$h$) free-surface synthetic Schlieren method. 

In this article, we limit this reconstruction to third order. Let us explain how to successively find $h^{(1)}$, $h^{(2)}$, and $h^{(3)}$. The reconstruction starts by assuming that $\delta \mathbf{r} = O (\varepsilon)$. The first-order problem is
\begin{equation}
    \delta\mathbf{r} = -\alpha H_{\mathrm{eff}} \boldsymbol{\nabla} h^{(1)},
    \label{eq:h1_sys}
\end{equation}
and recovers the linear (LNR) model of \citet{moisy2009synthetic}. The first-order elevation is obtained via integration:
\begin{equation}
    h^{(1)} = -\frac{1}{\alpha H_{\mathrm{eff}}} \boldsymbol{\nabla}^{-1}(\delta\mathbf{r}) + C^{(1)},
    \label{eq:h1}
\end{equation}
where $C^{(1)}$ is an integration constant and $\boldsymbol{\nabla}^{-1}$ denotes the inverse gradient operator. In practice, this gradient inversion problem is solved numerically using a spatial discretization of the gradient (see \S\ref{subsubsec:G} for more details).
The second order problem ($\mathcal{O}(\varepsilon^2)$) requires:
\begin{equation}
    \mathbf{0} = -\alpha \boldsymbol{\nabla} \left( H_{\mathrm{eff}} h^{(2)} + \frac{1}{2}(h^{(1)})^2 \right).
    \label{eq:h2_sys}
\end{equation}
Integration is immediate and yields the second-order correction:
\begin{equation}
    h^{(2)} = -\frac{(h^{(1)})^2}{2H_{\mathrm{eff}}} + C^{(2)}.
    \label{eq:h2}
\end{equation}
The third-order problem ($\mathcal{O}(\varepsilon^3)$) is :
\begin{equation}
    \begin{split}
        \mathbf{0} = &-\alpha \boldsymbol{\nabla} \left( H_{\mathrm{eff}} h^{(3)} + h^{(1)}h^{(2)} \right) \\
        &+ \frac{1}{2}\alpha^2 \left[ \eta H_{\mathrm{eff}} - \alpha(1-\eta_b^2)\eta_b H_b \right] \|\boldsymbol{\nabla} h^{(1)}\|^2 \boldsymbol{\nabla} h^{(1)},
    \end{split}
    \label{eq:h3_sys}
\end{equation}
which, upon integration, provides the third-order correction:
\begin{equation}
    \begin{split}
        h^{(3)} = &-\frac{h^{(1)}h^{(2)}}{H_{\mathrm{eff}}} - \frac{\eta}{2\alpha^2 H_{\mathrm{eff}}^3} \boldsymbol{\nabla}^{-1} \left( \|\delta \mathbf{r}\|^2 \delta \mathbf{r} \right) \\
        &+ \frac{\eta_b (1-\eta_b^2) H_b}{2\alpha H_{\mathrm{eff}}^4} \boldsymbol{\nabla}^{-1} \left( \|\delta \mathbf{r}\|^2 \delta \mathbf{r} \right) + C^{(3)}.
    \end{split}
    \label{eq:h3}
\end{equation}
Here we need one extra gradient inversion. The integration constants $C^{(i)}$ that appear at each step need to be determined using a physical boundary condition. In practice, we need to know the exact height of at least one point in the domain (e.g., pinned edges at the container walls). The absolute elevation is finally constructed as the sum of the successive components and we call $h \approx \sum_{i=1}^n h^{(i)}$, the $n$-th order WNL-$h$ reconstruction of the surface.

To estimate the convergence and error in this weakly nonlinear reconstruction, we consider a monochromatic surface wave with characteristic amplitude $|h| \sim A$ and wavenumber $k$. The small slope approximation requires $kA \ll 1$ and the small amplitude approximation requires $A/H_{\mathrm{eff}} \ll 1$. We define the leading-order elevation $h^{(1)} \sim \mathcal{O}(A)$ and estimate  $\boldsymbol{\nabla} \sim \mathcal{O}(k)$ and $\boldsymbol{\nabla}^{-1} \sim \mathcal{O}(k^{-1})$. We can then estimate that $\delta\mathbf{r}  \sim \ \mathcal{O}(\alpha H_{\mathrm{eff}} k A)$ and by induction, that
\begin{equation}
\begin{split}
h^{(2)} \sim & \  \mathcal{O}\left(\frac{A^2}{H_{\mathrm{eff}}}\right) \\
h^{(3)} \sim & \  \mathcal{O}\left(\frac{A^3}{H_{\mathrm{eff}}^2}\right) + \mathcal{O}\left(k^2 A^3\right) \\
h^{(4)} \sim & \ \mathcal{O}\left(\frac{A^4}{H_{\mathrm{eff}}^3}\right) + \mathcal{O}\left(\frac{k A^4}{H_{\mathrm{eff}}^2}\right) \\
    &+ \mathcal{O}\left(\frac{k^2 A^4}{H_{\mathrm{eff}}}\right) + \mathcal{O}\left(k^3 A^4\right). 
\end{split}
\end{equation}
Here  $h^{(4)}$ represents the fourth order correction that is not calculated but easily estimated. The relative reconstruction error for an $n$-th order approximation $h \approx \sum_{i=1}^n h^{(i)}$ is defined as
\begin{equation}
E^{(n)} = \frac{ \|h - \sum_{i=1}^n h^{(i)} \|_2 }{ \|h\|_2},
\end{equation}
where $\|...\|_2$ is the $L_2$-norm over the domain. We can estimate that:
\begin{equation}
\begin{split}
     E^{(1)} \sim & \ \mathcal{O}\left(\frac{A}{H_{\mathrm{eff}}}\right) ,\\
    E^{(2)} \sim & \ \mathcal{O}\left(\frac{A^2}{H_{\mathrm{eff}}^2}\right) + \mathcal{O}(k^2 A^2), \\
    E^{(3)} \sim & \ \mathcal{O}\left(\frac{A^3}{H_{\mathrm{eff}}^3}\right) + \mathcal{O}\left(\frac{k A^3}{H_{\mathrm{eff}}^2}\right) \\
    &+ \mathcal{O}\left(\frac{k^2 A^3}{H_{\mathrm{eff}}}\right) + \mathcal{O}(k^3 A^3). 
\end{split}
   \label{eq:error3}
\end{equation}
The dominant term in the relative error at second and third order depends on the value of $k H_{\mathrm{eff}}$. 

\subsubsection{WNL-$H$ algorithm \label{subsubsec:WNLH}}

When the wave amplitude $|h|$ is not small compared to the effective reference height $H_{\mathrm{eff}}$, the small-amplitude assumption ($h / H_{\mathrm{eff}}\sim\mathcal{O}(\varepsilon)$) made in the previous section is no longer valid. Errors will become very large in the WNL-$h$ reconstruction, see Eq.~\eqref{eq:error3}. We propose an alternative method that applies in the small slope limit but admits arbitrary amplitudes, $|h|/H_{\mathrm{eff}}\lesssim 1$.

Instead of calculating the elevation $h$, we directly find the instantaneous total height $H = H_{\mathrm{eff}} + h$. Since $\boldsymbol{\nabla} H = \boldsymbol{\nabla} h$ and $H\boldsymbol{\nabla} H = \boldsymbol{\nabla} (H^2/2)$, Eq.~(\ref{eq:dr_h}) can be rewritten as:
\begin{equation}
    \begin{split}
        \delta \mathbf{r}=&-\frac{\alpha}{2} \left(1-\frac{1}{2}\eta\alpha \|\boldsymbol{\nabla} H\|^2\right)\boldsymbol{\nabla} H^2\\
        &+\frac{1}{2}\alpha^3 (1-\eta_b^2)\eta_b H_b \|\boldsymbol{\nabla} H\|^2 \boldsymbol{\nabla} H.
    \end{split}
    \label{eq:dr_macthcalH}
\end{equation}
Given $\delta \mathbf{r}$, we can iteratively find a solution for the total squared height $H^2$. We denote $H_j$ the $j$-th estimate of the total effective height, with $j$ the iteration index. Considering that the slope is supposed small, we can find an initial guess $H_0^2$ via:
\begin{equation}
    H_0^2 = -\frac{2}{\alpha} \boldsymbol{\nabla}^{-1}(\delta \mathbf{r}) + C_0,
    \label{eq:initial_guess}
\end{equation}
where $C_0$ is an integration constant determined by boundary conditions. This reconstruction is the equivalent of what the method of \citet{li2021single} would give, without paraxial corrections. To reach higher precision,  we rearrange the full expression of Eq.~(\ref{eq:dr_macthcalH}) into an iterative scheme, where the $(j+1)$-th estimate is found by integration of :
\begin{equation}
    \boldsymbol{\nabla}(H_{j+1}^2) = -\frac{2}{\alpha}\frac{\delta \mathbf{r} - \frac{1}{2}\alpha^3 (1-\eta_b^2)\eta_b H_b \|\boldsymbol{\nabla} H_j\|^2 \boldsymbol{\nabla} H_j}{1 - \frac{1}{2}\eta\alpha \|\boldsymbol{\nabla} H_j\|^2 }.
    \label{eq:iteration_WNLH}
\end{equation}
At each iteration, an integration constant $C_{j+1}$ has to be set by a boundary condition (e.g., $H^2 = H_{\mathrm{eff}}^2$ at pinned boundaries). The process repeats until convergence, defined by $\|H_{j+1} - H_j\|_{2,\mathrm{max}} < e$, with $e$ the selected error tolerance. We denote $H_{\text{WNL}}$ the converged total effective height. Considering that Eq.~\eqref{eq:dr_macthcalH} is only valid for small slopes, we call this the weakly nonlinear WNL-$H$ method. 

This nonlinear reconstruction of the total effective height $H$ only requires small slopes $kA \ll 1$, but there are no restrictions on the ratio $A/H_{\mathrm{eff}}$. The relative error on the initial guess $H_0$ and converged state $H_{\text{WNL}}$ are expected to scale as 
\begin{equation}
 \begin{split}
E_0 = & \frac{ \| H - H_0  \|_2}{\|H\|_2}  \sim \mathcal{O} ( (kA)^2 ) \\
E_{\text{WNL}} = & \frac{ \| H - H_{\text{WNL}}  \|_2}{\|H\|_2}  \sim \mathcal{O}( (kA)^4 ).
 \end{split}
 \label{eq:error_WNLH}
\end{equation}

The iterative scheme in Eq.~(\ref{eq:iteration_WNLH}) provides a very simple way to solve the nonlinear problem \eqref{eq:dr_macthcalH}. There is no guarantee that it always converges but practice shows that it works well. 

\subsubsection{FNL-$H$ algorithm  \label{subsubsec:FNLH}}

The iterative logic of the WNL-$H$ approach can be extended to solve for the effective height $H$ using the exact refraction model. Hence it is possible to relax both assumptions of small slopes and small amplitudes. 

To explain the iterative algorithm, we first rewrite the liquid-induced and bottom-induced displacements in a condensed way as:
\begin{equation}
\begin{split}
    \delta\mathbf{r}_l=& -\mathcal{A}(\boldsymbol{\nabla} H)(H_l+h)\boldsymbol{\nabla} H, \\
    \delta\mathbf{r}_b=& -\mathcal{B}(\boldsymbol{\nabla} H)\eta_b H_b\boldsymbol{\nabla} H.
\end{split}
\end{equation}
Here $\mathcal{A}(\boldsymbol{\nabla} H)$ and $\mathcal{B}(\boldsymbol{\nabla} H)$ are fully nonlinear functions of $\boldsymbol{\nabla} H$ that are given in Eq.~(\ref{eq:dr_l}) and Eq.~(\ref{eq:dr_b}), respectively. The total displacement is expressed as:
\begin{equation}
\delta\mathbf{r} = -\mathcal{A}(\boldsymbol{\nabla} H)H\boldsymbol{\nabla} H + [\mathcal{A}(\boldsymbol{\nabla} H)-\mathcal{B}(\boldsymbol{\nabla} H)]\eta_b H_b\boldsymbol{\nabla} H.
\end{equation}

Given $\delta \mathbf{r}$, we iteratively solve for the total squared height $H^2$.  We use the same initial guess $H_0$ of Eq.~(\ref{eq:initial_guess}) and then we iterate on the following scheme:
\begin{equation}
    \frac{\boldsymbol{\nabla}(H_{j+1}^2)}{2} = -\frac{\delta\mathbf{r} - [\mathcal{A}(\boldsymbol{\nabla} H_j)-\mathcal{B}(\boldsymbol{\nabla} H_j)]\eta_b H_b \boldsymbol{\nabla} H_j}{\mathcal{A}(\boldsymbol{\nabla} H_j)}.
    \label{eq:iteration_WNLH_full}
\end{equation}
In the specific case where no bottom spacer is present ($H_b = 0$), the iterative scheme simplifies significantly:
\begin{equation}
H_{j+1}^2 = -\frac{2}{\mathcal{A}(\boldsymbol{\nabla} H_j)}\boldsymbol{\nabla}^{-1}(\delta\mathbf{r}) + C_{j+1},
\label{eq:iteration_WNLH_Hb0}
\end{equation} 
In our tests, we find that this algorithm converges well in a few iterations. We denote the converged solution $H_{\text{FNL}}$. This fully nonlinear approach is termed as the FNL-$H$ method and theoretically, it can recover the exact surface shape, for vanishing noise, without spatial discretization errors and as long as there are no caustics. As in the WNL-$H$ scheme, there is no guarantee for convergence, but practice demonstrates that the FNL-$H$ scheme is highly robust and effective.

\section{Numerical implementation\label{subsec:workflow}\label{sec:practical}}

\begin{figure*}
    \centering
    \includegraphics{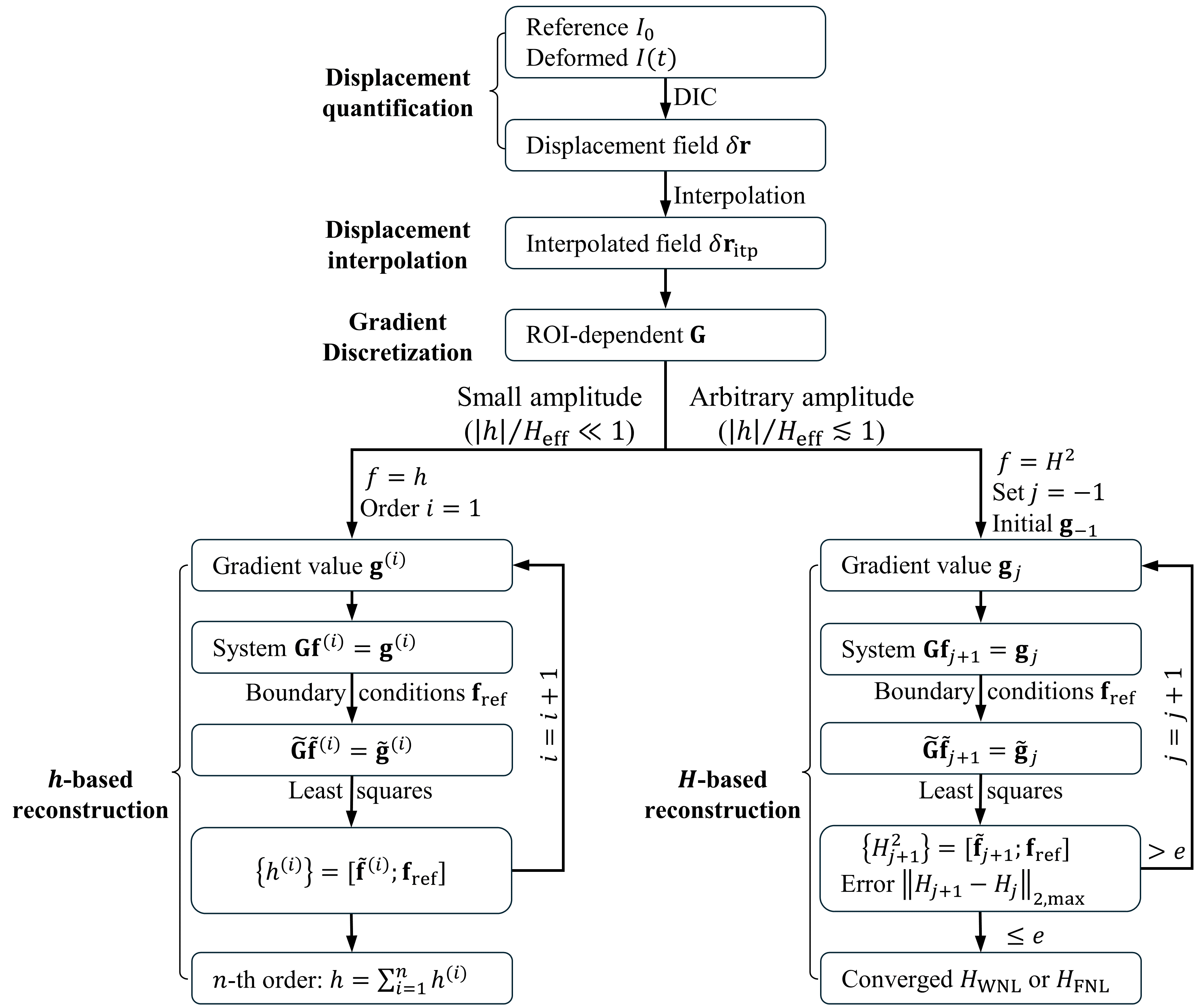}
    \caption{Workflow charts of the $h$-based (WNL-$h$) method and the $H$-based (WNL-$H$ or FNL-$H$) method}
    \label{fig:workflow}
\end{figure*}
The practical implementation of the nonlinear surface reconstruction algorithms follows a multi-stage workflow that is illustrated in Fig.~\ref{fig:workflow} and discussed here. 

\subsection{Displacement quantification}

In a typical experimental setup, the liquid container, the reference dot pattern, and the camera system are initially leveled under quiescent conditions. A reference image $I_0$ of a random-dot pattern is acquired in the quiescent state (flat interface), followed by a refracted image $I_1$ (or a time-series of refracted images $I(t)$).  We introduce a Cartesian reference frame $(O,\hat{\mathbf{x}},\hat{\mathbf{y}})$ aligned with the camera pixel array (Fig.~\ref{fig:schematicMapping}). We denote $\mathbf{r} = \mathbf{OP}$ the coordinates of the points on the quiescent reference pattern and $\mathbf{r}' = \mathbf{OP}'$ represent its displaced virtual image. The image is divided into a grid of interrogation windows of size $\Delta L$, and the local displacement field $\delta\mathbf{r}  = \mathbf{r}'  - \mathbf{r} $ is computed by correlating $I_1$ (or $I(t)$) with $I_0$ in each interrogation window via a DIC (digital image correlation) algorithm.  The displacement field $\delta \mathbf{r} $ is therefore defined on a regular, square $\mathbf{r}$-grid as in Fig.~\ref{fig:schematicMapping}(a) (blue circles), with a spatial resolution $\Delta L$ determined by the size of the interrogation window. The design of the random-dot pattern is critical to ensure high accuracy in the measurement of the displacement field  (experimental details are provided in section 5). 

\subsection{Displacement interpolation\label{sec:itp}}
\begin{figure}
    \centering
    \includegraphics{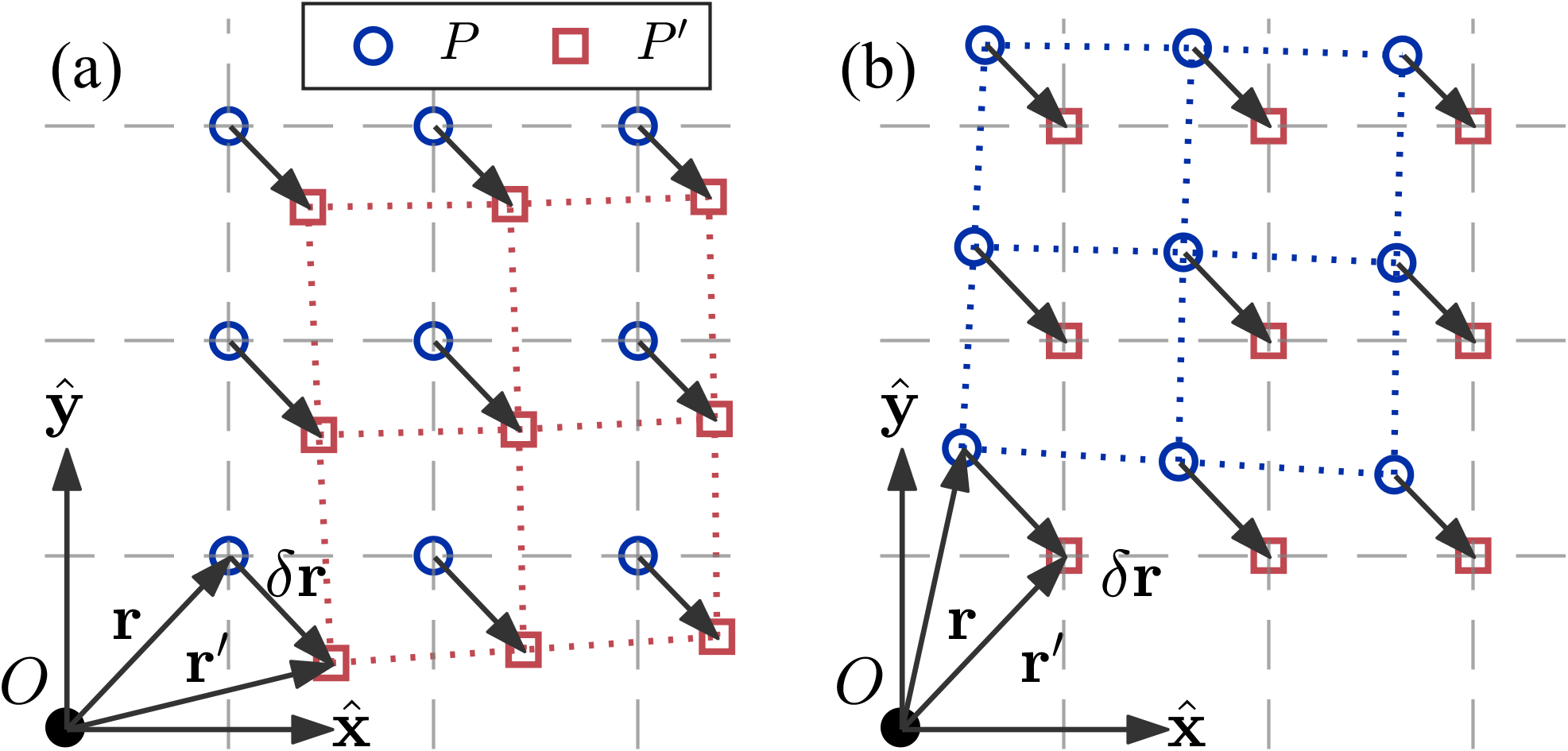}
    \caption{Schematic illustration of the reference grid $P$ (blue circles), the deformed grid $P'$ (red squares), and the resulting displacement field $\delta\mathbf{r} = \mathbf{r}' - \mathbf{r}$. (a) Raw displacement field measured via DIC. (b) Interpolated displacement field after coordinate mapping. The vectors $\mathbf{r}$ and $\mathbf{r}'$ denote the positions of points $P$ and $P'$, respectively, within the Cartesian coordinate system}
    \label{fig:schematicMapping}
\end{figure}

The theoretical refraction equations found in the previous section relate the displacement $\delta \mathbf{r}$ at position $\mathbf{r}$, to  $\boldsymbol{\nabla} h$ and $h$ at position $\mathbf{r}'$  (Fig.~\ref{fig:light_refraction}).  This difference in location creates a supplementary difficulty: the raw displacement field  $\delta \mathbf{r}$ is measured on a {\it regular} $\mathbf{r}$-grid, but relates to the gradient of elevation on an {\it irregular} $\mathbf{r}'$-grid. Although gradient inversion can, in principle, be done on irregular grids, it is more convenient to use regular grids, in particular when handling image sequences.


The simple strategy to handle this difficulty is to calculate an interpolated displacement field  $\delta\mathbf{r}_{\text{itp}}$ so that the associated $\mathbf{r}'$-grid where we want to invert the gradient is {\it regular} and {\it time-independent}.  This is  illustrated in Fig.~\ref{fig:schematicMapping}(b):  the interpolated displacement field $\delta\mathbf{r}_{\text{itp}}$ links non-uniformly distributed initial positions $\mathbf{r}$ (blue circles) to uniformly distributed $\mathbf{r}'$-points (red squares). 

In practice, this interpolation is done using standard libraries in Matlab (\texttt{interp2}) or Python (\texttt{scipy.interpolate}). The precision depends on the grid-size $\Delta L$ and the chosen interpolation scheme, typically linear, cubic or spline. The difference between the raw and interpolated displacement is of second order in wave-magnitude and for this reason, interpolation is only necessary in nonlinear reconstructions algorithms. In the original, linear reconstruction algorithm of \citet{moisy2009synthetic}, interpolation does not increase precision.

\subsection{Iterative surface reconstruction}


\subsubsection{Discretising \& inverting gradients\label{subsubsec:G}}

All the reconstruction methods discussed above rely on solving a sequence of over-determined problems of the form $\boldsymbol{\boldsymbol{\nabla}} f = {\bf g}$. Here $f$ represents $h$ or $H^2$ and ${\bf g}$ is the measured gradient. In the simplest applications, the regions of interest (ROI) is rectangular; however, it is often useful to work within non-rectangular ROI, which may include masked regions (for instance, in the presence of floating objects).

\begin{figure}
    \centering
    \includegraphics{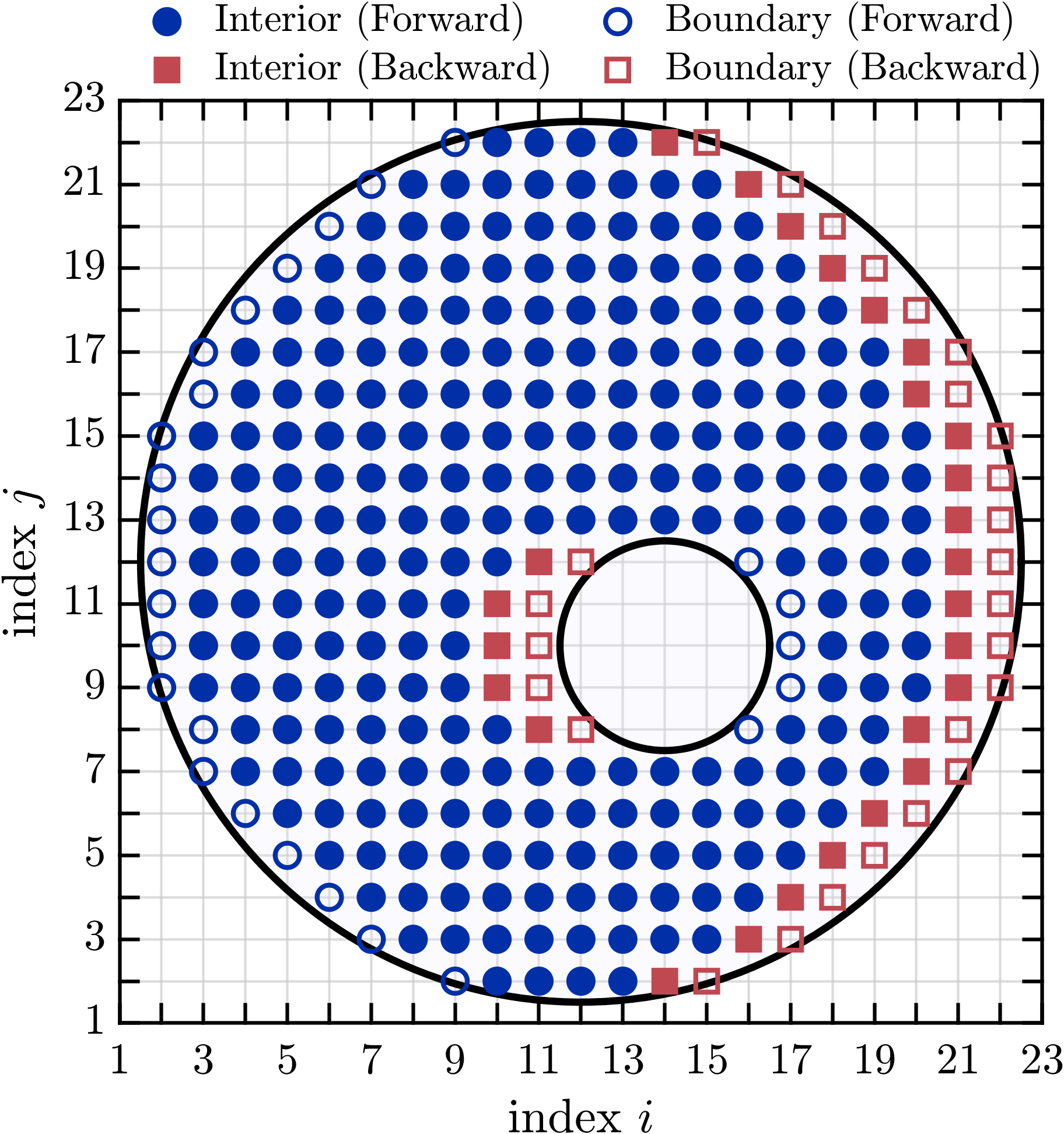}
    \caption{Schematic illustration of the adaptive discretization for the partial derivative  $\partial/ \partial x$ within an arbitrary ROI. Full and hollow symbols denote interior and near-boundary nodes, respectively. Circles and squares indicate nodes utilizing forward and backward difference schemes, respectively}
    \label{fig:ROI}
\end{figure}

In our implementation, we use second or third order finite differences defined on an arbitrary ROI. Consider for instance the ROI of Fig.~\ref{fig:ROI}, delimited by the outer and inner circles, with an initial square data-set for the interpolated displacement defined on a $23 \times 23$ uniform grid with cell-size $\Delta L$. The discretization of the gradient on points inside the ROI is done using 4 different schemes, depending on where the point is located. Let $f_{i,j}$ denote the field value at position $i,j$ in the ROI, and consider first the horizontal derivative $\partial f / \partial x$. On all {\it interior} positions marked as full blue circles, we use a third-order, four-point forward scheme:
\begin{equation}
    \left. \frac{\partial f}{\partial x} \right|_{i,j} \approx \frac{-2f_{i-1,j} - 3f_{i,j} + 6f_{i+1,j} - f_{i+2,j}}{6\Delta L}.
\end{equation}
Interior points, marked as full red squares, have the neighboring $(i+2,j)$ point outside the ROI. At these positions, we rather use a third-order, four-point backward scheme:
\begin{equation}
    \left. \frac{\partial f}{\partial x} \right|_{i,j} \approx \frac{f_{i-2,j} - 6f_{i-1,j} + 3f_{i,j} + 2f_{i+1,j}}{6\Delta L}.
\end{equation}
At the immediate boundaries of the ROI, left or right neigbors $(i\pm1,j)$ do not exist. The discretization scheme is there changed to a second-order, three-point forward discretization (hollow blue circles):
\begin{equation}
    \left. \frac{\partial f}{\partial x} \right|_{i,j} \approx \frac{-3f_{i,j} + 4f_{i+1,j} - f_{i+2,j}}{2\Delta L},
\end{equation}
or to a second-order, three-point backward discretization (hollow red squares):
\begin{equation}
    \left. \frac{\partial f}{\partial x} \right|_{i,j} \approx \frac{f_{i-2,j} - 4f_{i-1,j} + 3f_{i,j}}{2\Delta L}.
\end{equation}
The partial derivatives $\partial f / \partial y$ are discretized analogously. In cases where local obstructions or narrow gaps result in fewer than three available nodes, our framework adaptively reverts to a first-order scheme to maintain numerical stability. Furthermore, for configurations where the exact boundary location is known with high precision, more sophisticated boundary-conforming treatments can be incorporated, as demonstrated by \citet{zhang2023pattern}. This adaptive strategy ensures that the gradient operator remains well-defined even for highly non-convex domains. 

Applying this discretization strategy to all $N$ points inside the ROI, we get a discretised version of the problem $\boldsymbol{\boldsymbol{\nabla}} f = \mathbf{g}$ that takes the shape of an overdetermined linear system: 
\begin{equation}
    \mathbf{G} \mathbf{f} = \mathbf{g}. 
    \label{eq:linear_system}
\end{equation}
Here $\mathbf{f} = \{f_{i,j}\}$ is the $N \times 1$ column-vector of the unknown nodal values of $f$ and  $\mathbf{g} =  \{{g_{x}}_{i,j}, {g_{y}}_{i,j} \}$ is the $2 N \times 1$ column-vector of known gradient values. The sparse  $2 N \times N$ matrix $\mathbf{G}$ represents the discretised gradient operator and it depends exclusively on the fixed mesh geometry and ROI. The matrix $\mathbf{G}$ remains constant throughout the iterative reconstruction process and is calculated once and stored. It is also used in the WNL-$H$ and FNL-$H$ methods to compute gradients via a single sparse matrix-vector multiplication. 

To obtain a unique solution $h$ or $H^2$ and resolve the integration constant $C^{(i)}$ or $C_j$, the system (\ref{eq:linear_system}) is partitioned into unknown values and fixed boundary constraints. Let $\mathbf{f}_{\mathrm{ref}}$ represent the vector of $N_{\mathrm{ref}}$ prescribed reference values, and $\tilde{\mathbf{f}}$ denote the vector of the remaining $N - N_{\mathrm{ref}}$ unknown nodal values. The system is subsequently reformulated as:
\begin{equation}
    \tilde{\mathbf{G}} \tilde{\mathbf{f}} = \tilde{\mathbf{g}} = \mathbf{g} - \mathbf{G}_{\mathrm{ref}} \mathbf{f}_{\mathrm{ref}},
    \label{eq:partitioned_system}
\end{equation}
where $\tilde{\mathbf{G}} \in \mathbb{R}^{2N \times (N - N_{\mathrm{ref}})}$ and $\mathbf{G}_{\mathrm{ref}} \in \mathbb{R}^{2N \times N_{\mathrm{ref}}}$ are sub-matrices of $\mathbf{G}$ composed of the columns corresponding to the unknown and boundary nodes, respectively. The term $\mathbf{G}_{\mathrm{ref}} \mathbf{f}_{\mathrm{ref}}$ accounts for the contribution of the fixed boundary heights to the gradient field $\tilde{\mathbf{g}}$.

The resulting over-determined system (\ref{eq:partitioned_system}) is solved in a least-squares sense, e.g. using the backslash operator ($\tilde{\mathbf{f}} = \tilde{\mathbf{G}} \backslash \tilde{\mathbf{g}}$) in MATLAB or via \texttt{scipy.sparse.linalg} in Python (refer to  \url{https://github.com/zhang-shimin/NLSS} for implementation details). The resulting solution $\mathbf{f} = [\tilde{\mathbf{f}}; \mathbf{f}_{\mathrm{ref}}]$ provides the reconstructed surface elevation or total effective height at all nodes in ROI.

\subsubsection{Update loops}

Everything is now in place to iteratively reconstruct the surface.  The reconstruction paths vary depending on the chosen algorithms and in all formula, we need to use the interpolated displacement $\delta\mathbf{r}_{\text{itp}}$: 
\begin{itemize}
    \item {WNL-$h$}: The first-order elevation $h^{(1)}$ is obtained by solving the linear system $\boldsymbol{\nabla} h^{(1)} = -\delta\mathbf{r}_{\text{itp}}/(\alpha H_{\mathrm{eff}})$ with boundary condition $h^{(1)}(\mathbf{r}_{\text{ref}}) = h_{\text{ref}}$. The second-order correction $h^{(2)}$ is found from Eq.~\ref{eq:h2} with boundary condition  $h^{(2)}(\mathbf{r}_{\text{ref}}) =0$. For the third-order correction $h^{(3)}$ (Eq.~\ref{eq:h3}) we need one extra gradient inversion and use the boundary condition $h^{(3)}(\mathbf{r}_{\text{ref}}) =0$. The final WNL $n$th-order topography is the cumulative sum $h \approx \sum_{i=1}^n h^{(i)}$. Each inversion is computed as a least-squares solution to the system $\tilde{\mathbf{G}} \tilde{\mathbf{f}}^{(i)} = \tilde{\mathbf{g}}^{(i)}$ (Fig.~\ref{fig:workflow}, left loop).
    \\
    \item {WNL-$H$ and FNL-$H$}: Both algorithms are initialized with the initial guess $H_0^2$ by solving $\tilde{\mathbf{G}} \tilde{\mathbf{f}}_{0} = \tilde{\mathbf{g}}_{-1}$, using the linear approximation $\mathbf{g}_{-1} = -(2/\alpha)\{\delta\mathbf{r}_{\text{itp}}\}$ (see Eq.~\ref{eq:initial_guess}). Starting from this guess, the algorithm determines the update $H_{j+1}^2$ by inverting the gradient $\boldsymbol{\nabla}(H_{j+1}^2)$ via Eq.~(\ref{eq:iteration_WNLH}) for WNL-$H$ or Eq.~(\ref{eq:iteration_WNLH_full}) for FNL-$H$. Each update is performed via the same least-squares framework $\tilde{\mathbf{G}} \tilde{\mathbf{f}}_{j+1} = \tilde{\mathbf{g}}_{j}$ (Fig.~\ref{fig:workflow}, right loop), with the gradient $\boldsymbol{\nabla} H_j$ numerically evaluated using the sparse matrix $\mathbf{G}$. The boundary constraint $H_j^2(\mathbf{r}_{\text{ref}}) = H_{\mathrm{eff}}^2$ is enforced at every iteration. The process terminates upon reaching a converged $H_{\text{WNL}}$ or $H_{\text{FNL}}$ once the condition $\|H_{j+1}-H_j\|_{2,\text{max}} < e$ is satisfied.
\end{itemize}

\section{Numerical tests\label{sec:numTest}}

To validate the proposed reconstruction methods, we first conduct a series of numerical tests in an idealized, noise-free configuration. We simulate an air-water interface ($n_a = 1$ and $n_l = 1.33$) within an acrylic container ($n_b = 1.49$). The  surface topography is prescribed as:
\begin{equation}
    h_{\mathrm{exact}}(x,y) = A \sin(kx) \sin(ky),
    \label{eq:h_exact}
\end{equation}
where $A$ is the wave amplitude and $k = 2\pi/\lambda$ represents the wavenumber for a characteristic wavelength $\lambda = 20\ \mathrm{mm}$. The computational domain covers a square region of interest (ROI) $[0, \lambda] \times [0, \lambda]$, discretized into a regular grid with a spatial resolution of $ \Delta L = 0.5\ \mathrm{mm}$.  In this numerical test, we directly compute the exact displacement field $\delta\mathbf{r}_*$ using Eqs.~(\ref{eq:dr_l}) and (\ref{eq:dr_b}) on a uniform $\mathbf{r}'$-grid (see Fig. \ref{fig:schematicMapping}b), so we skip the interpolation step here. This provides an exact synthetic dataset that allows a direct evaluation of the reconstruction accuracy independently of noise and interpolation errors.

To assess the operational limits of the WNL and FNL algorithms, we vary the amplitude $A$ and the reference height $H_{\mathrm{eff}}$, thereby exploring a broad range of wave steepness ($kA$) and amplitude-to-depth ratios ($A/H_{\mathrm{eff}}$). However, in a real setup, part of the parameter space remains inaccessible: for one-dimensional sinusoidal waves, the condition $A/H_{\mathrm{eff}} \gtrsim \alpha (kA)^{2}$ \citep{moisy2009synthetic} leads to ray crossing and caustic formation, beyond which the displacement fields can no longer be reliably measured. The present numerical test does not suffer from this caustic problem as we calculate the displacement field analytically, rather than measuring it from a refracted image. It is, however, possible to identify the regions of parameter space where caustics are expected to occur by detecting negative local gradients in the deformed grid coordinates, which indicate a loss of monotonicity.

In all reconstructions, we set the origin as the reference point, $\mathbf{r}_\text{ref} = \mathbf{0}$ and we fix the absolute height. In the WNL-$h$ method, we have $h^{(i)}(0,0) = 0$ for all $i$ and for the iterative $H$-based methods, we have $H_j^2(0,0) = H_{\mathrm{eff}}^2$. For the WNL-$H$ and FNL-$H$ schemes, the solution is considered converged when the maximum change in the reconstructed depth between successive iterations falls below a specified tolerance, $\| H_j - H_{j-1} \|_{2,\text{max}} < e$, where we set $e = 10^{-7}\ \mathrm{mm}$. It should be noted that an algorithmic breakdown of the $H$-based methods can occur in regimes with extreme surface slopes or shallow depths. Specifically, if the iterative update for the squared depth $H_j^2$ becomes negative, the solution enters a non-physical domain where $H_j$ cannot be resolved, at which point the iteration is terminated, and the solution from the previous stable step is taken as the final reconstruction.

To evaluate and compare the performance of the different frameworks, we define the following reconstructed elevations and their corresponding errors:
\begin{itemize}
    \item $ \sum_{i=1}^n h^{(i)}$ and $E^{(n)}$ 
 \end{itemize}
for the $n$-th order WNL-$h$ method, and     
\begin{itemize}
    \item $h_0 = H_0 - H_{\mathrm{eff}}$ and $E_0$ 
    \item $h_{\mathrm{WNL}} = H_{\mathrm{WNL}} - H_{\mathrm{eff}}$ and $E_{\mathrm{WNL}}$
    \item $h_{\mathrm{FNL}} = H_{\mathrm{FNL}} - H_{\mathrm{eff}}$ and $E_{\mathrm{FNL}}$ 
\end{itemize}
for the initial guess and converged solutions in the WNL-$H$ and FNL-$H$ methods. The reconstruction error, $E$, is calculated using the relative $L_2$-norm by substituting the respective reconstructed elevation, $h_{\mathrm{test}}$, into the following expression:
\begin{equation}
    E = \frac{\sqrt{\sum_{i,j} |h_{\mathrm{exact}}(x_i, y_j) - h_{\mathrm{test}}(x_i, y_j)|^2}}{\sqrt{\sum_{i,j} |h_{\mathrm{exact}}(x_i, y_j)|^2}}.
    \label{eq:Er}
\end{equation}


\subsection{Performance of WNL-$h$ method}

\begin{figure}
    \centering
    \includegraphics[width=\linewidth]{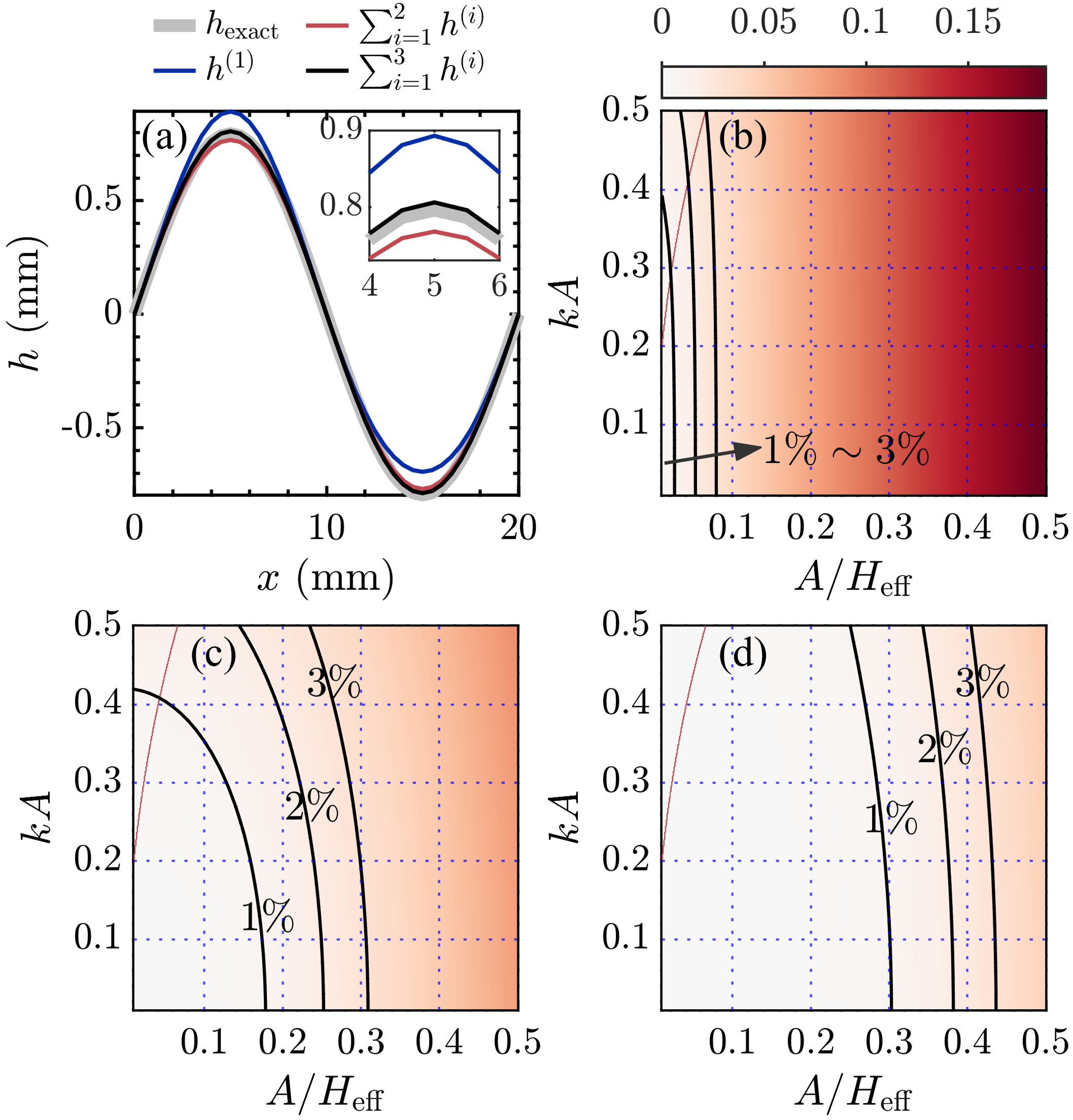}
    \caption{Numerical validation of the WNL-$h$ reconstruction algorithm. (a) Comparison of reconstructed surface profiles at $y = 5\ \mathrm{mm}$ against the prescribed exact elevation $h_{\mathrm{exact}}$ under $kA=A/H_{\mathrm{eff}}=0.25$ for successive orders $n=1,2,3$; the inset provides a magnified view of the peak region to highlight the convergence toward the exact solution. Relative error distributions ($E^{(n)}$) are shown for the (b) first-order (LNR), (c) second-order, and (d) third-order WNL methods. The solid red curve denotes the simulated boundary for caustic formation (region to the left). Three black contours indicate the $1\%\sim 3\%$ relative error thresholds for each method}
    \label{fig:numTest_WNL}
\end{figure}

We first evaluate the performance of the WNL-$h$ method in a set-up without a solid layer ($H_b = 0$). We vary wave steepness $kA \in (0, 0.5]$ and relative amplitude-to-depth ratio $A/H_{\mathrm{eff}} \in (0, 0.5]$.  

In Fig.~\ref{fig:numTest_WNL}(a) we compare the exact wave profile and the reconstructed profiles with the sum truncated at order 1, 2 and 3, in the case $kA= A/H_{\mathrm{eff}} = 0.25$. This figure clearly demonstrates the increased accuracy when incorporating higher-order corrections, with the third order reconstruction nearly capturing the exact profile. There are notable differences with the first order linear reconstruction at the peak values.

In Fig.~\ref{fig:numTest_WNL}(b$\sim$d), we compare the relative errors $E^{(1)}$,  $E^{(2)}$ and  $E^{(3)}$ of first, second and third order reconstructions  in the $kA$, $A/H_{\mathrm{eff}}$ plane. We clearly see that relative errors decrease as the reconstruction order $n$ increases. Caustics would prevent reconstructions in the small part of parameter space left of the red line.

\begin{figure}
    \centering
    \includegraphics{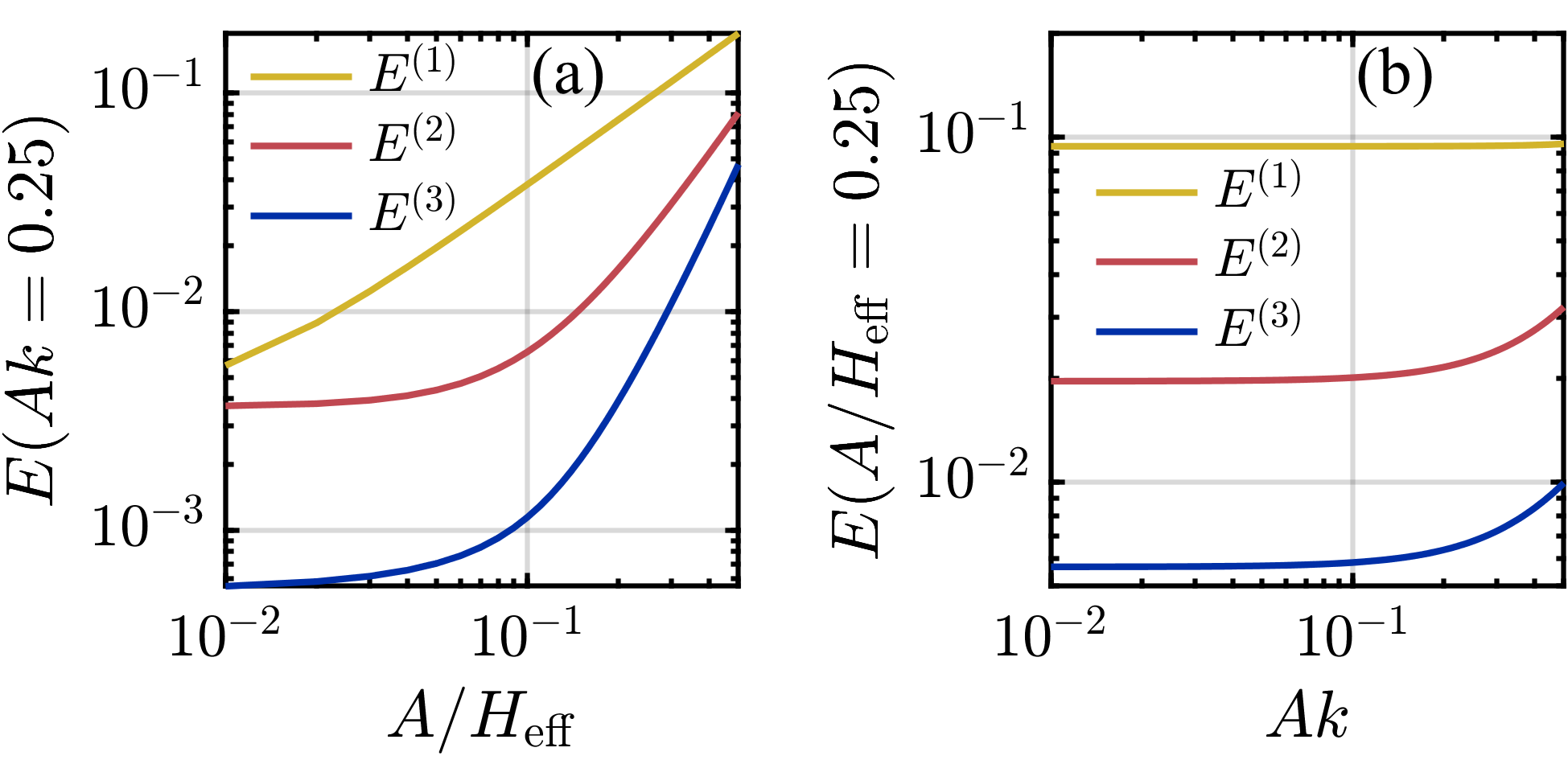}
    \caption{Relative reconstruction errors $E^{(n)}$ for the $n$-th order WNL-$h$ methods ($n=1,2,3$), evaluated under (a) $Ak = 0.25$ and (b) $A/H_{\mathrm{eff}} = 0.25$}
    \label{fig:numTest_WNL_profile}
\end{figure}

To test whether the error scales as predicted by theory,  Eq.~\eqref{eq:error3},  we extract two characteristic error curves at fixed $kA = 0.25$ and at fixed $A/H_{\mathrm{eff}} = 0.25$, as shown in Fig.~\ref{fig:numTest_WNL_profile}(a) and (b), respectively. In Fig.~\ref{fig:numTest_WNL_profile}(a), the relative error follows the predicted power-law scaling $E^{(n)} \sim \mathcal{O}((A/H_{\mathrm{eff}})^n)$, consistent with the theoretical error analysis. It should be noted that this scaling law reaches a floor at very small $A/H_{\mathrm{eff}}$ values, where the third-order discretization errors of the gradient operator of order $\mathcal{O}(\Delta L^3)$ become the dominant error source. Conversely, the power-law dependence on $kA$ is not apparent in Fig.~\ref{fig:numTest_WNL_profile}(b) within the explored parameter range. This is because the errors associated with the wave steepness remain smaller than those governed by the amplitude-to-depth ratio and gradient discretization. The contour topology in Fig.~\ref{fig:numTest_WNL} further supports this: in this test, the ratio $A/H_{\mathrm{eff}}$ is the primary source of reconstruction error in the WNL-$h$ framework. 

\subsection{Performance of $H$-based methods}
\label{sec:numTest_H_based}

\begin{figure}
    \centering
    \includegraphics[width=\linewidth]{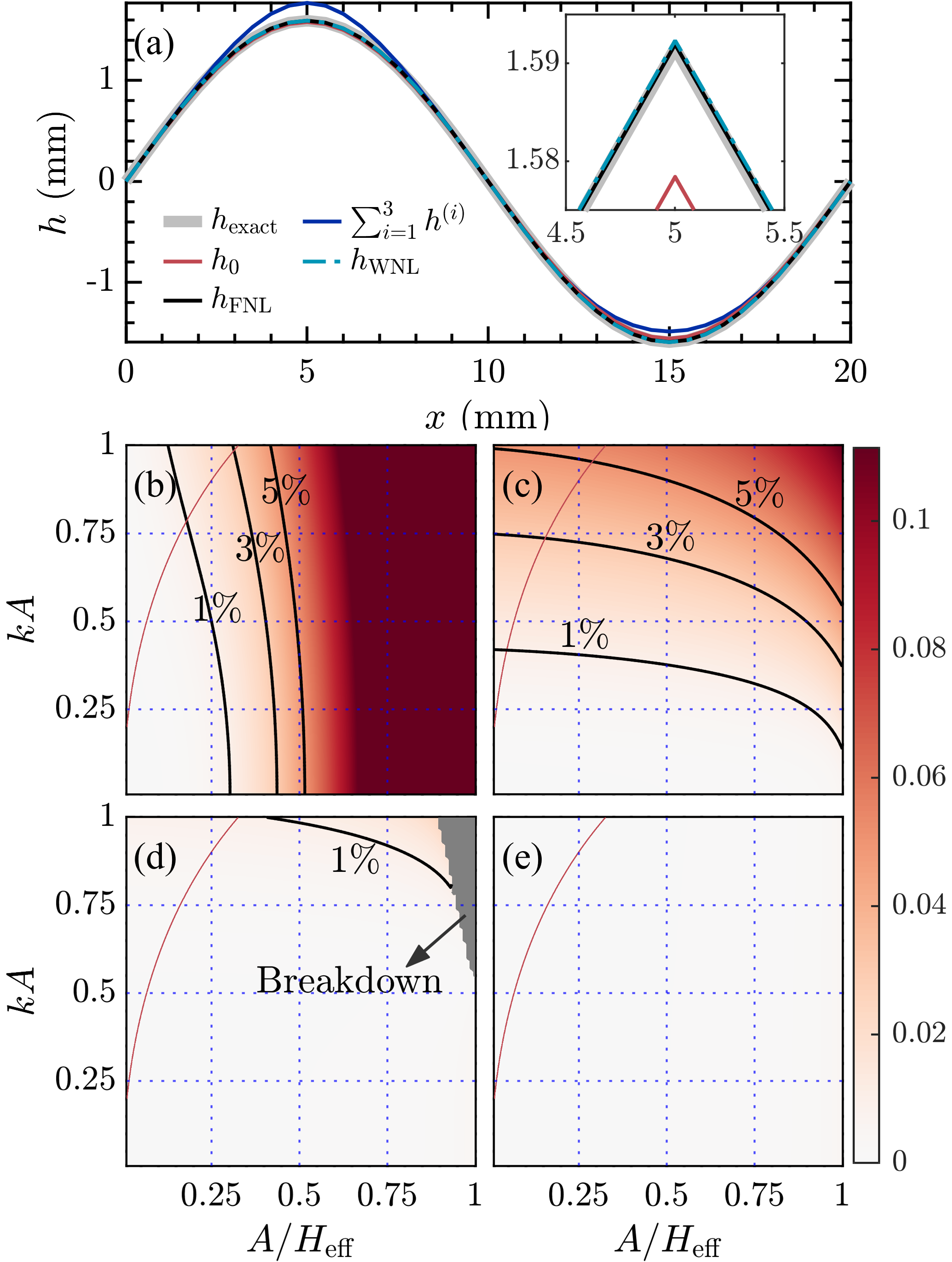}
    \caption{Comparison of reconstruction accuracy between the WNL-$h$ method and the iterative $H$-based methods (WNL-$H$ and FNL-$H$). (a) Reconstructed surface profiles at $y = 5\ \mathrm{mm}$ compared against the prescribed exact elevation $h_{\mathrm{exact}}$ for $kA=A/H_{\mathrm{eff}}=0.5$; the inset provides a magnified view of the peak region to highlight the convergence toward the exact solution. Relative error distributions ($E$) are displayed for: (b) the third-order WNL-$h$ method ($E^{(3)}$), (c) the initial guess ($E_0$), (d) the WNL-$H$ method ($E_{\text{WNL}}$), and (e) the FNL-$H$ method ($E_{\text{FNL}}$).  The caustic boundary (solid red curve) and relative error contours (black curves with indications) are superimposed to illustrate the expanded valid measurement range. A distinct dashed region in (d) identifies the breakdown zone where the WNL-$H$ iterative scheme fails to converge}
    \label{fig:numTest_FNL}
\end{figure}

\begin{figure}
    \centering
    \includegraphics{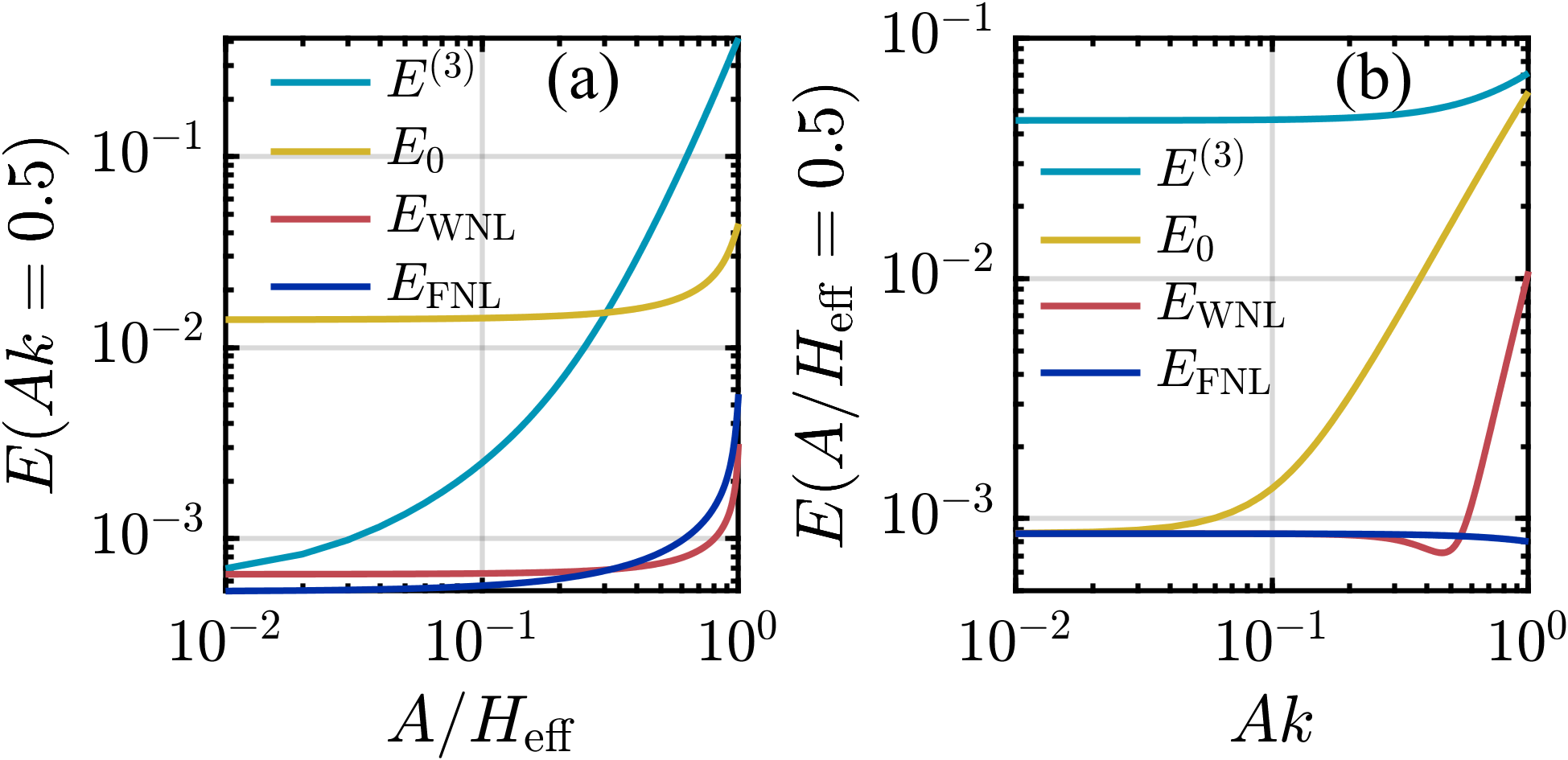}
    \caption{Comparison of relative reconstruction errors $E$ between the $h$-based method (WNL-$h$) and the $H$-based iterative methods (WNL-$H$ and FNL-$H$). Performance is evaluated under (a) $Ak = 0.5$ and (b) $A/H_{\mathrm{eff}} = 0.5$}
    \label{fig:numTest_FNL_profile}
\end{figure}

We now consider the two $H$-based iterative methods: the weakly nonlinear WNL-$H$ and the fully nonlinear FNL-$H$. Their performance, in comparison with the WNL-$h$ method, is illustrated in Fig.~\ref{fig:numTest_FNL}. For this purpose, we explore an extended parameter range, $A/H_{\mathrm{eff}} \in (0, 1]$ and $kA \in (0, 1]$, well beyond the linear regime.

In Fig.~\ref{fig:numTest_FNL}(a) we see for $kA= A/H_{\mathrm{eff}} = 0.5$, that the reconstructed elevation profiles are accurate, but that even the third-order WNL-$h$ reconstruction starts to exhibit significant deviations from the exact profile, particularly at the wave crests and troughs. In contrast, the $H$-based methods demonstrate superior performance. This is further reflected in the error maps (see Figs.~\ref{fig:numTest_FNL}b-e), where we see that the $H$-based approaches drastically reduce the error levels observed in the WNL-$h$ framework. While the WNL-$h$ method is fundamentally limited by the amplitude-to-depth ratio, the $H$-based schemes can still successfully reconstruct the surface even for high values of $A/H_{\mathrm{eff}}$. We noticed that for the highest slopes, the iterative WNL-$H$ scheme does not always converge, see Fig.~\ref{fig:numTest_FNL}(d). The FNL-$H$ algorithm performs best, Fig.~\ref{fig:numTest_FNL}(e), with a small error that is only due to spatial discretization. 

The scaling properties of the error are further analyzed in Fig.~\ref{fig:numTest_FNL_profile}. For a fixed wave steepness $kA = 0.5$ (see Fig.~\ref{fig:numTest_FNL_profile}a), the WNL-$h$ error follows the predicted power law $E^{(3)} \sim \mathcal{O}((A/H_{\mathrm{eff}})^3)$, as derived in Eq.~(\ref{eq:error3}). Conversely, the $H$-based methods (WNL-$H$ and FNL-$H$) exhibit an error floor nearly independent of $A/H_{\mathrm{eff}}$ as expected by theory. When the amplitude ratio is fixed at $A/H_{\mathrm{eff}} = 0.5$ (see Fig.~\ref{fig:numTest_FNL_profile}b), the error for the initial guess scales as $E_0 \sim \mathcal{O}((kA)^2)$ and that for the WNL-$H$ method scales as $E_{\text{WNL}} \sim \mathcal{O}((kA)^4)$, matching the theoretical expectation in Eq.~(\ref{eq:error_WNLH}). The residual error in the FNL-$H$ method is nearly constant as it is governed by spatial discretization errors.

\begin{figure}
    \centering
    \includegraphics{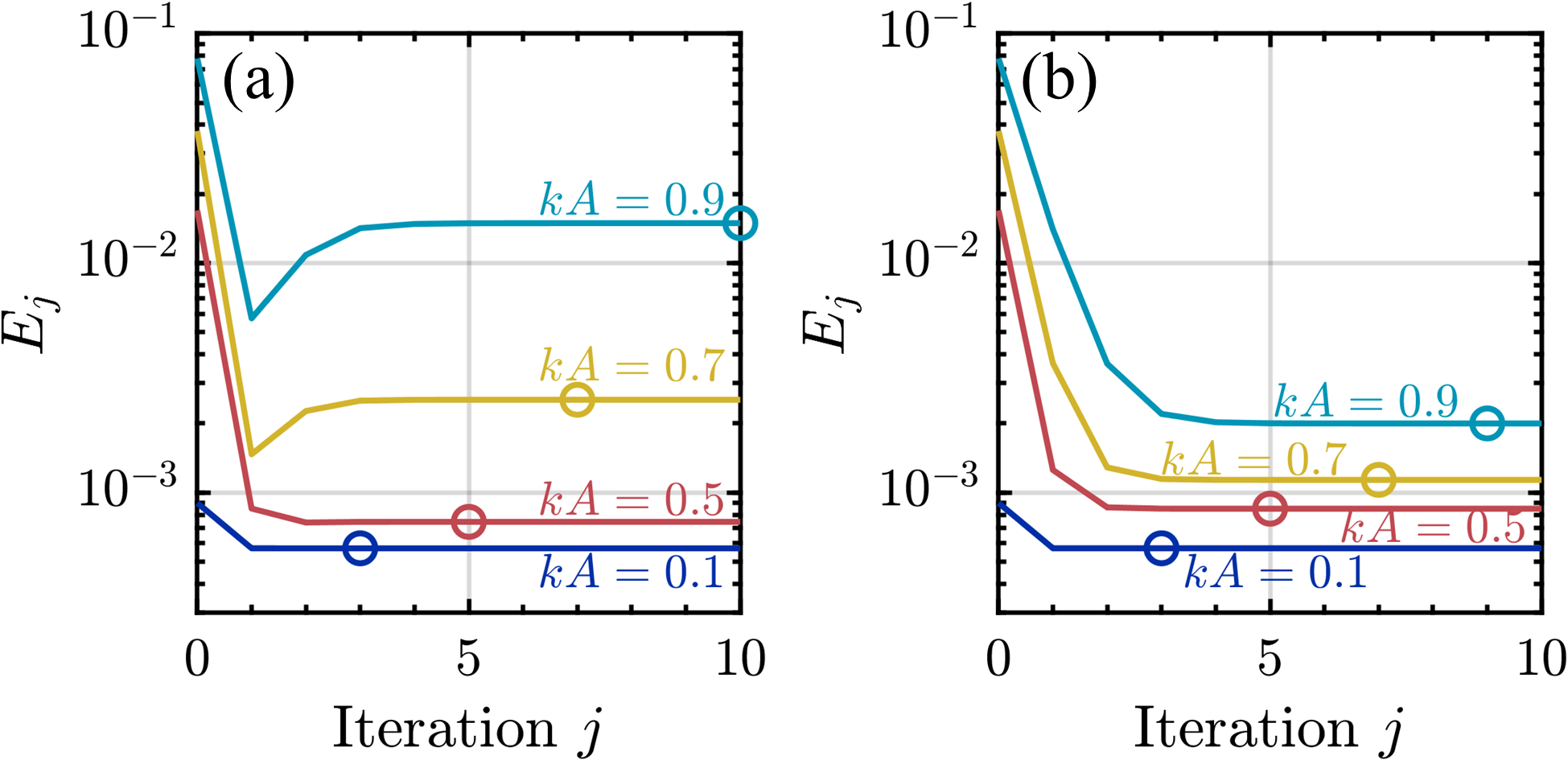}
    \caption{Iteration-wise error convergence of the $H$-based frameworks. The relative reconstruction error $E_j$ is plotted against the iteration number $j$ for the (a) WNL-$H$ and (b) FNL-$H$ methods across varying regimes of $kA = A/H_{\mathrm{eff}} = 0.1, 0.5, 0.7, \text{and } 0.9$. The convergence points are highlighted with circles for each curve}
    \label{fig:RE_interation}
\end{figure}

In Fig.~\ref{fig:RE_interation}, we show how the reconstruction error decreases as a function of iteration number in the WNL-$H$ and FNL-$H$ update loops. We typically need no more than 5 iterations before reaching a plateau or fixed point in the iterative scheme, which demonstrates the computational effectiveness of the WNL-$H$ and FNL-$H$ methods. For the WNL-$H$ method (Fig.~\ref{fig:RE_interation}a), the reconstruction error $E_j$ decreases monotonically through the iterations at low $kA$ but not for high $kA$. This is not unexpected, as we are leaving the small-slope approximation.
The FNL-$H$ method (Fig.~\ref{fig:RE_interation}b) on the other hand always exhibits a consistent monotonic reduction in error across all tested regimes.

\subsection{Effect of secondary layer thickness}

\begin{figure}
    \centering
    \includegraphics{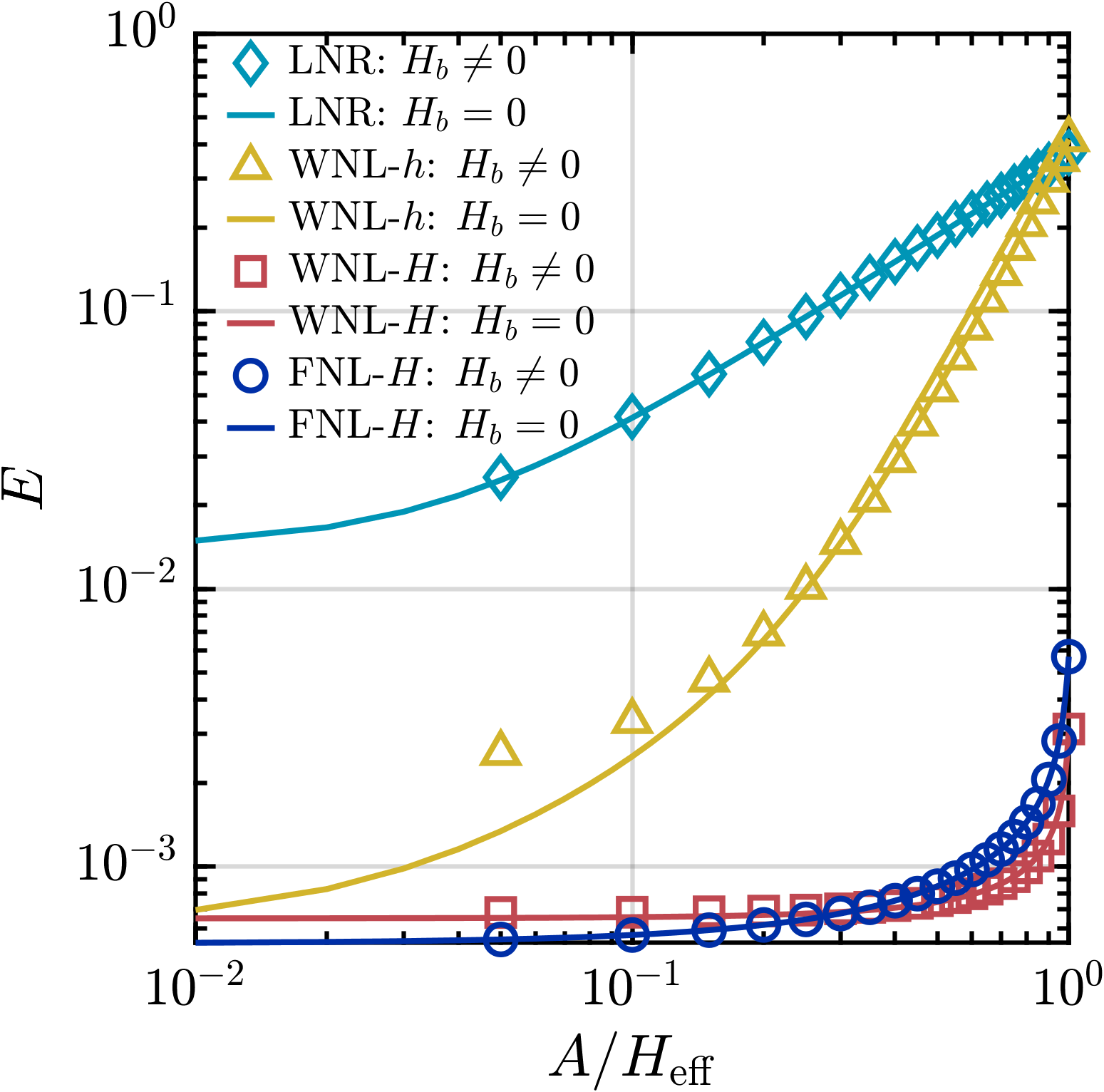}
    \caption{Influence of the secondary layer thickness $H_b$ on the reconstruction error $E$ for the LNR, WNL-$h$, WNL-$H$, and FNL-$H$ methods under $kA=0.5$ and $A/H_{\mathrm{eff}}=1$. As $H_b$ increases, the effective height-to-depth ratio $A/H_{\mathrm{eff}}$ decreases. For comparison, the corresponding solid error curves (representing $H_b = 0$) are extracted from the previous results}
    \label{fig:RError_Sigma}
\end{figure}
In most realistic set-ups, we have a bottom spacer ($H_b\neq 0$), and this increases the total effective depth $H_{\mathrm{eff}}$ in the surface reconstruction. Increasing $H_b$ therefore is a simple way to reduce the effective amplitude-to-depth ratio $A/H_{\mathrm{eff}}$ and its associated error. This is visible in Fig.~\ref{fig:RError_Sigma}. The symbols illustrate the error variation for fixed parameters $kA=0.5$ and $A/H_l = 1$, while $A/H_{\mathrm{eff}}$ is varied within the range $[0.05, 1]$ by adjusting $H_b\geq 0$. The full lines, show error curves from the previous tests with $H_b=0$, $kA = 0.5$ and varying $A/H_l$. Curves with and without spacers nearly coincide, confirming that $A/H_{\mathrm{eff}} $ is indeed the parameter that controls error.

\begin{figure}
    \centering
    \includegraphics{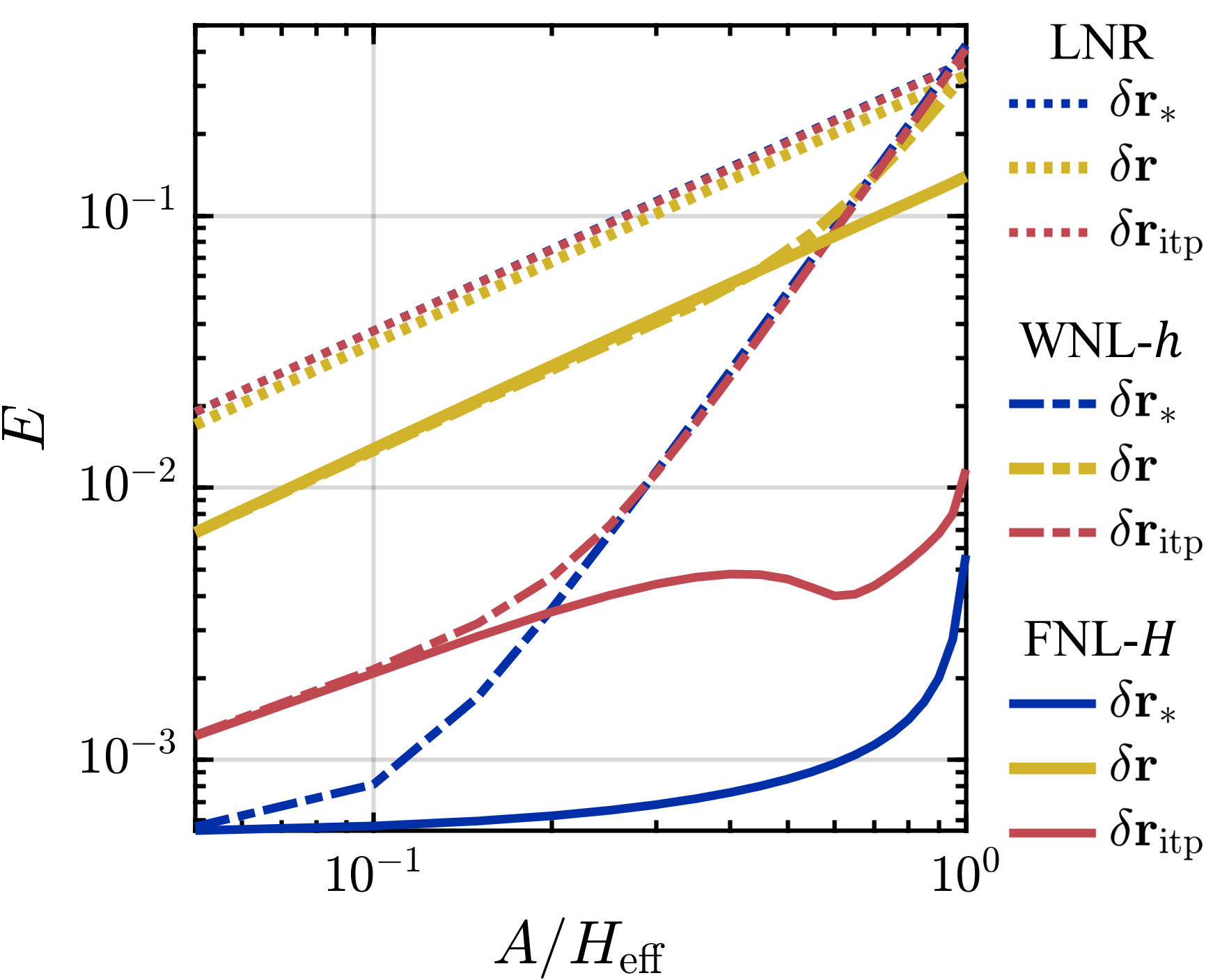}
    \caption{Impact of displacement interpolation on reconstruction accuracy. Relative reconstruction error $E$ for the LNR, WNL-$h$, and FNL-$H$ methods, comparing the performance when using different displacement fields (raw  $\delta\mathbf{r}$, interpolated $\delta\mathbf{r}_{\mathrm{itp}}$ and ideal $\delta\mathbf{r}_*$) with fixed $H_{\mathrm{eff}} = 10/\pi\ \mathrm{mm}$ ($H_b=0$) and varying $A$}
    \label{fig:numTest_mapped}
\end{figure}

\subsection{Influence of interpolation}

As explained in  \S \ref{sec:itp}, it is necessary to interpolate the measured displacement field onto the regular reference grid. In this section, we demonstrate the importance of this interpolation step in the nonlinear surface reconstructions and we quantify the interpolation error.  For this, we take $H_b = 0$, fix $H_{\text{eff}} = H_l= 1/k = 10/\pi\ \mathrm{mm}$ and vary $A/H_{\text{eff}} \in [0.05, 1]$ by adjusting $A$. Referring to Fig.~\ref{fig:schematicMapping}, we calculate three different displacement fields: 

\begin{itemize}
\item $\delta\mathbf{r}_*$: ``ideal'' displacement from an irregular $\mathbf{r}$-grid to the regular $\mathbf{r}'$-grid, as in Fig.~\ref{fig:schematicMapping}(b). As in the previous sections, we can calculate this displacement field directly from the refraction formula, using the nodal values of $h$ and $\boldsymbol{\nabla} h$ on the regular $\mathbf{r}'$-grid. 

\item $\delta\mathbf{r}$: ``raw'' displacement from the regular $\mathbf{r}$-grid to an irregular $\mathbf{r}'$-grid, as illustrated in Fig.~\ref{fig:schematicMapping}(a). This raw displacement mimics DIC measurements, i.e., it reproduces the apparent shift of a reference pattern viewed through the interface. For this, we resolve the ray-path for each point $P$ on the regular $\mathbf{r}$-grid by numerically determining the intersection point $P_s$ at the free surface satisfying the refraction law. Specifically, we minimize the refraction residual $\|\eta \mathbf{n}_s \times \mathbf{e}_z - \mathbf{n}_s \times \mathbf{v}_i\|$, where $\mathbf{v}_i$ is the unit vector of the incident ray $\mathbf{PP}_s$ given by $\mathbf{v}_i = \mathbf{PP}_s/ ||\mathbf{PP}_s|| $, with $\mathbf{PP}_s =\mathbf{r}' - \mathbf{r} + (H + h(\mathbf{r}')) \hat{\mathbf{z}}$. This nonlinear system is solved using a standard iterative solver (e.g., \texttt{fsolve} in MATLAB).


\item $\delta\mathbf{r}_{\mathrm{itp}}$: ``interpolated'' displacement from an irregular $\mathbf{r}$ grid to the regular $\mathbf{r}'$ grid, as in Fig.~\ref{fig:schematicMapping}(b). From the exact displacement data $\delta\mathbf{r}$ and associated  irregular $\mathbf{r}'$-grid, we interpolate a displacement field $\delta\mathbf{r}_{\mathrm{itp}}$ associated with the regular $\mathbf{r}'$-grid via a cubic interpolation scheme. This interpolated field approaches that of $\delta\mathbf{r}_*$ and differences are only due to interpolation error. \\

\end{itemize}
With these data sets we can compare different surface reconstructions.  In a first series of ``ideal" reconstructions, we use the exact displacement field $\delta\mathbf{r}_*$ to reconstruct the surface, just as in the previous sections. The reconstruction errors ($E$) of the LNR, third-order WNL-$h$ and FNL-$H$ reconstructions as a function of $A/H_\text{eff}$ are shown as blue curves in Fig.~\ref{fig:numTest_mapped}. We observe the expected trends, with error growing linearly ($ E \sim A/H_{\text{eff}}$)  in the LNR-reconstruction, cubically ($ E \sim (A/H_{\text{eff}})^3$) with third order WNL-$h$. Error is minimal when we use the FNL-$H$ method.  

In the second series, we compute the surface with the ``raw'' field $\delta\mathbf{r}$ which mimics the data typically produced by DIC. By doing so, we deliberately reproduce the error that would arise from considering the measured field on the non-interpolated grid. The corresponding errors are shown by the yellow curves in Fig.~\ref{fig:numTest_mapped} that show linear trends ($ E \sim A/H_{\text{eff}}$) for all methods. This demonstrates that high-order accuracy cannot be reached with the nonlinear methods if we use the raw displacement field $\delta\mathbf{r}$. Only for the LNR method, the distinction between $\delta\mathbf{r}$ and $\delta\mathbf{r}_*$ is not important. 

In the third series, we use the ``interpolated'' displacement  field $\delta\mathbf{r}_{\mathrm{itp}}$ to reconstruct the surface. Although interpolation introduces supplementary errors, it aims to correct the error generated by the geometric shift between the reference grid on the dot pattern and the refracted points on the free interface.  As shown by the red curves in Fig.~\ref{fig:numTest_mapped}, the reconstruction error of the WNL-$h$ and FNL-$H$ reconstructions is indeed below the yellow curves, showing that interpolation effectively reduces  error. The difference between the red (interpolated) and blue (best possible) curves corresponds to the interpolation error and depends on the used interpolation scheme and resolution $\Delta L$. 

This test shows that interpolation is not merely an optimization of the nonlinear reconstruction but a prerequisite for reaching high order precision: we really need to calculate $\delta\mathbf{r}_{\mathrm{itp}}$ before applying the nonlinear reconstruction algorithms in the experimental implementations. 

\section{Experimental tests\label{sec:expTest}}

We have applied the three nonlinear reconstruction methods to three distinct experiments. First, we benchmark the algorithms against a known ground-truth geometry using a precision-machined plano-convex glass lens. In a second set-up, we reconstruct the air-liquid interfaces of a slowly spreading silicone oil drop on a glass substrate. In a third set-up, we  measure  the time-dependent surface elevation in periodic Faraday waves.

In all configurations, a high-resolution ($2592\times 2048\ \mathrm{pixel}^2$) camera equipped with a 0.125X telecentric lens (TechSpech GoldTL 2/3'') is used to capture the reference and refracted dot patterns. The telecentric lens constrains the effective field of view to a circular region approximately $90$~mm in diameter. The reference pattern consists of randomly distributed $0.2$-$\mathrm{mm}$ black dots printed on a matte white background. The displacement fields are computed using the high-resolution DIC algorithm available in the open-source Matlab toolbox \textit{PIVlab} \citep{stamhuis2014pivlab}. We use a multi-pass fast Fourier transform (FFT) algorithm with interrogation windows of size iteratively reduced from $64 \times 64$ to $16 \times 16\ \mathrm{pixel}^2$ with a constant 50\% overlap, resulting in a final spatial resolution of $\Delta L=8\,\mathrm{pixel}=0.295$~mm. The dot images have a diameter of $5\sim 6$ pixels, with a density of $3\sim 4$ dots per $16 \times 16~\text{pixel}^2$ interrogation window, consistent with the requirements for high-precision DIC. Finally, the measured displacement fields are interpolated onto a regular grid using cubic interpolation, and subsequently processed through the various reconstruction frameworks {, with a tolerance $e = 10^{-7}\ \mathrm{mm}$ set for the $H$-based methods, }to enable a comparative performance analysis.

\subsection{Plano-convex glass lens}

\begin{figure}[htbp]
    \centering
    \includegraphics{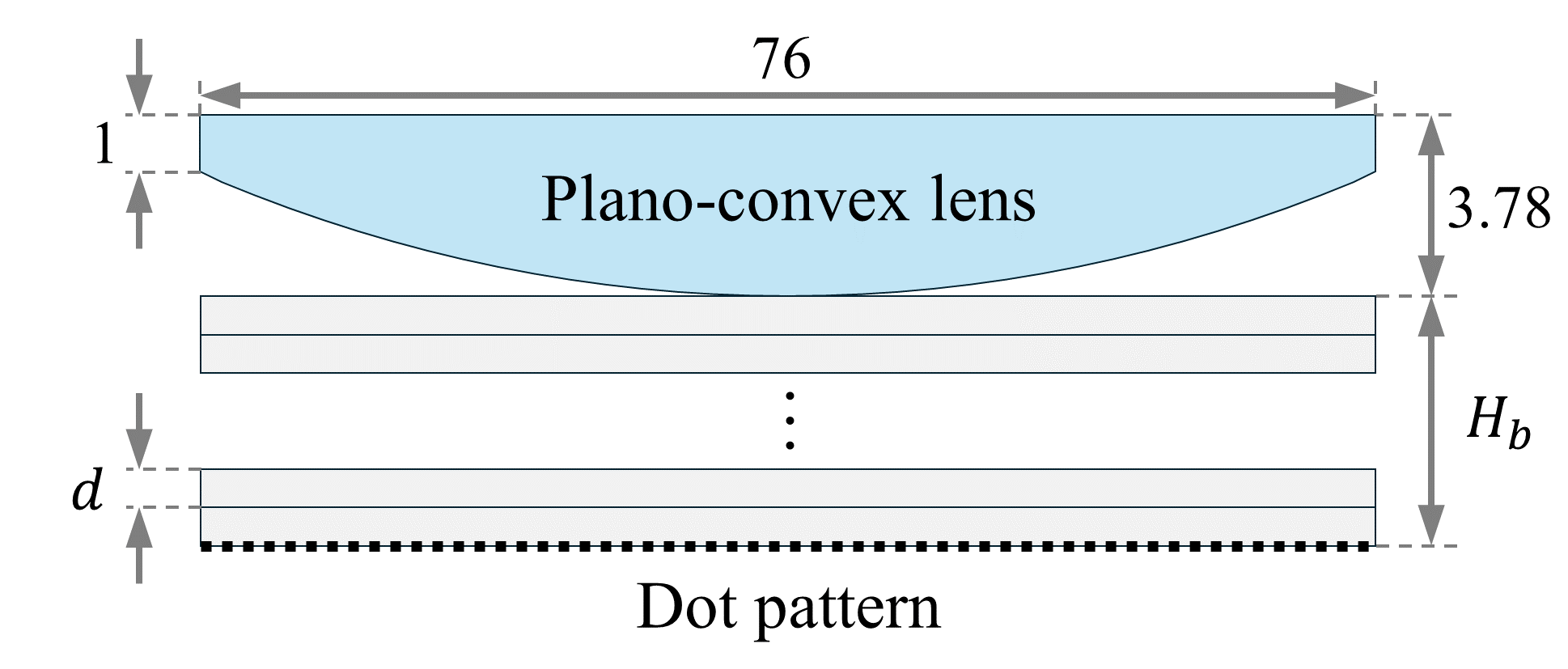}
    \caption{Experimental layout of the plano-convex glass lens assembly. $\mathcal{N}$ acrylic plates, each of thickness $d$, are stacked beneath the lens to modulate the effective depth, $H_{\mathrm{eff}} = H_b = \mathcal{N}d$}
    \label{fig:lensLayout}
\end{figure}

The plano-convex glass lens has a diameter of 76 mm, a vertex thickness of 3.78 mm, and an edge thickness of 1.00 mm. Its spherical profile, with a radius of curvature $R = 261.1$ mm, provides a long focal length of 503 mm, which ensures that ray crossing (caustics) is physically precluded. As illustrated in Fig.~\ref{fig:lensLayout}, the lens is oriented with its spherical interface, $h_{\mathrm{exact}}$, facing downward, yielding a refractive index ratio of $\eta \approx 1.52$ at the glass-air interface. To systematically modulate the ratio $h_{\mathrm{exact}}/H_{\mathrm{eff}}$, a stack of $\mathcal{N}$ flat acrylic plates (each with thickness $d = 2$~mm and refractive index $n_b \approx 1.49$) is positioned beneath the lens, resulting in an index ratio of $\eta_b = 1/n_b \approx 0.67$ at the air-acrylic interface. This setup provides a controlled base thickness $H_b = \mathcal{N}d$. The reference elevation is set to zero at the lens vertex: $h^{(i)}(r=0) = 0$ for the WNL-$h$ method and $H_j^2(r=0) = H_{\mathrm{eff}}^2 = H_b^2$ for the $H$-based methods. The resulting height fields are azimuthally averaged to produce radial profiles $h(r)$.

\begin{figure}[htbp]
    \centering
    \includegraphics{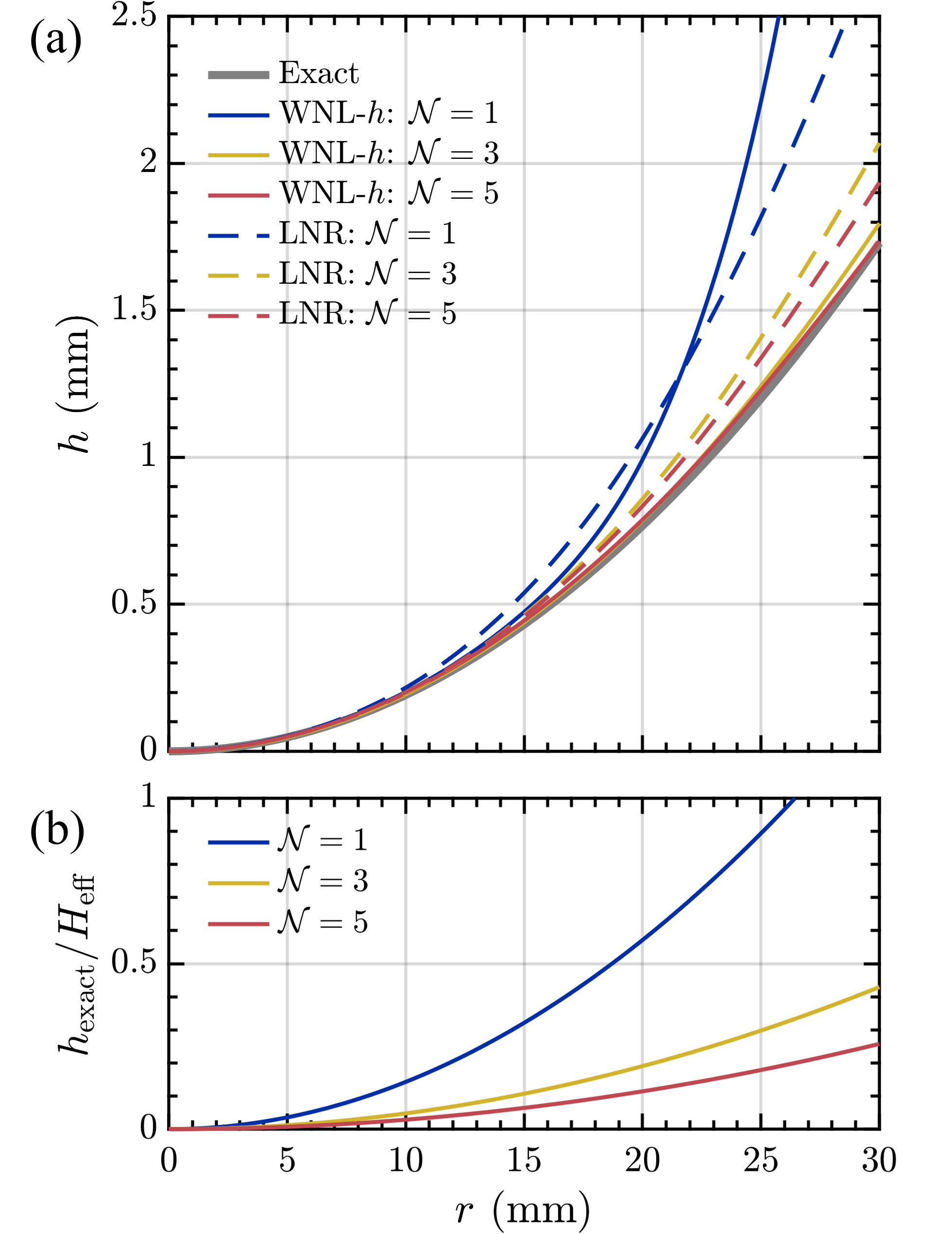}
    \caption{Reconstruction accuracy of the WNL-$h$ method for the plano-convex lens. (a) Comparison of surface profiles using first-order (LNR) and third-order (WNL-$h$) methods across varying plate counts $\mathcal{N}$. The exact lens profile (gray solid line) is determined from geometric specifications. (b) Corresponding spatial distribution of the local elevation-to-depth ratio $h_{\mathrm{exact}}/H_{\mathrm{eff}}$ along the lens radius}
    \label{fig:lensReconsProfle}
\end{figure}

The performance of the first-order (LNR) and third-order WNL-$h$ methods is shown in Fig.~\ref{fig:lensReconsProfle}(a). For both methods, increasing the number of plates $\mathcal{N}$ improves the reconstruction accuracy due to the significant reduction in the local elevation-to-depth ratio $h_{\mathrm{exact}}/H_{\mathrm{eff}}$, as depicted in Fig.~\ref{fig:lensReconsProfle}(b). Notably, for $\mathcal{N}=1$, the ratio exceeds 0.5 at large radii; in this regime, the third-order WNL-$h$ result exhibits an enhancement of the measurement error that eventually exceeds even the LNR result. This behavior aligns with our numerical predictions (see Sec.~\ref{sec:numTest_H_based}) and highlights the fundamental limitation of WNL-$h$ methods when the small-amplitude approximation is not satisfied.

\begin{figure}[htbp]
    \centering
    \includegraphics{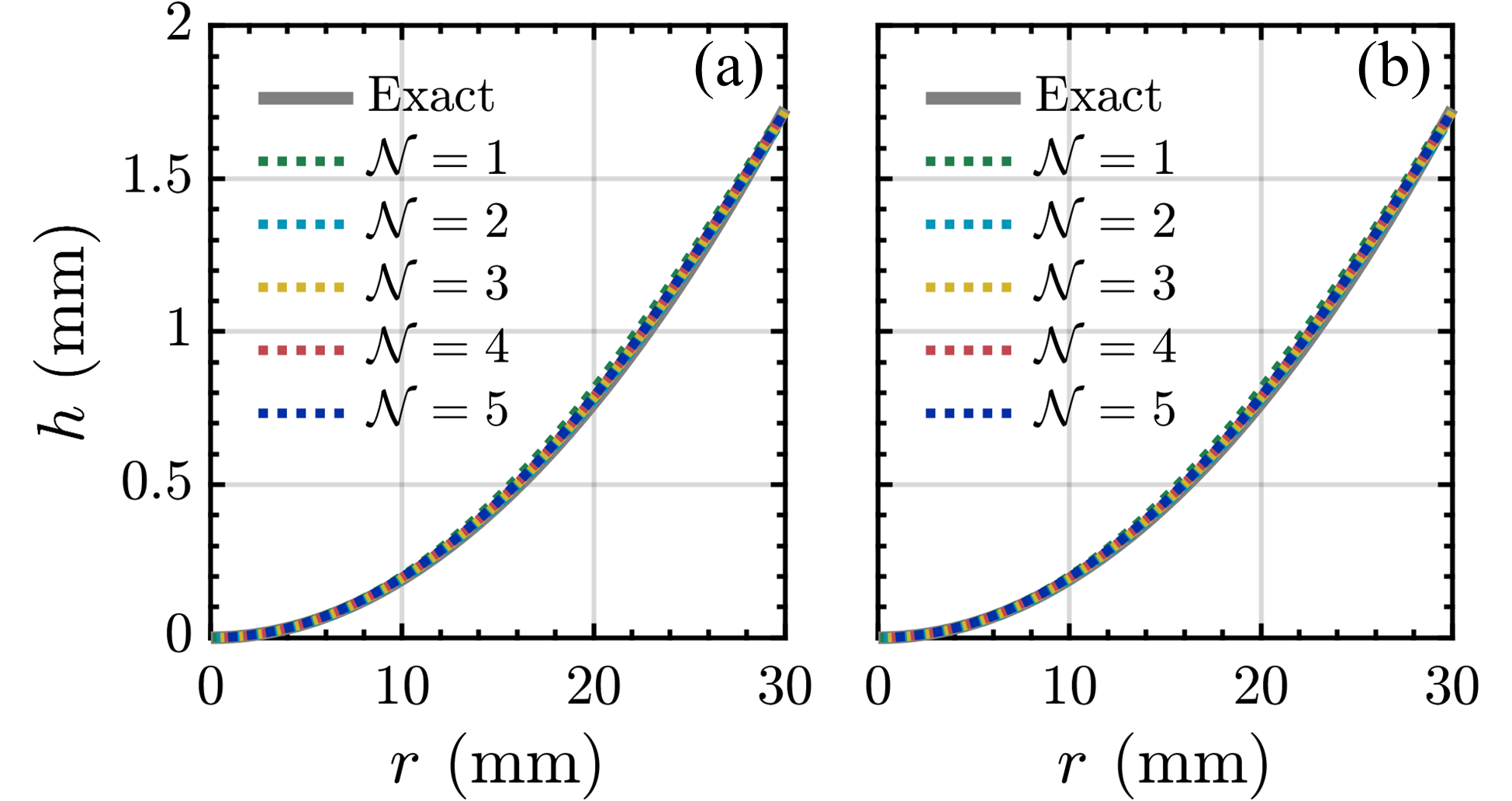}
    \caption{Reconstruction accuracy of the iterative $H$-based frameworks for the plano-convex lens. Surface profiles are obtained via the (a) WNL-$H$ and (b) FNL-$H$ methods across varying secondary layer depths $H_b = \mathcal{N}d$.}
    \label{fig:lensReconsProfle_FNL}
\end{figure}

In contrast, the iterative $H$-based methods achieve high-accuracy reconstructions even at large $h_{\mathrm{exact}}/H_{\mathrm{eff}}$ ratios, as illustrated in Fig.~\ref{fig:lensReconsProfle_FNL}. Both the WNL-$H$ and FNL-$H$ methods demonstrate excellent performance that is effectively independent of the elevation-to-depth ratio across all tested values of $\mathcal{N}$. Due to the large focal length of the lens, the maximum gradient $\|\boldsymbol{\nabla} h\|$ remains small and no significant discrepancy is observed between the WNL-$H$ and FNL-$H$ methods.


\subsection{Drop spreading on a solid substrate}

To evaluate the algorithms using liquids under quasi-static conditions, we investigate the spreading dynamics of a silicone oil drop (viscosity $100\ \mathrm{mPa\cdot s}$ and volume $\sim 12$~mm$^3$) deposited on a flat glass substrate ($H_b = 1\ \mathrm{mm}$). The refractive index ratios are set to $\eta = 1/1.40$ (air-oil) and $\eta_b = 1.40/1.52$ (oil-glass). Here, the ROI includes the drop and the dry part of the plate, so no mean elevation $H_l$ can be defined. The effective depth is defined as $H_{\mathrm{eff}} = \eta_b H_b$. To rigorously evaluate the algorithms in this high-amplitude regime, the glass thickness was intentionally chosen to be comparable to the initial drop height ($h/H_{\mathrm{eff}} \approx 0.8$). Images were acquired at 10~Hz, with the reference height set to zero within the dry domain far from the contact line.

\begin{figure}[htbp]
    \centering
    \includegraphics[width=\linewidth]{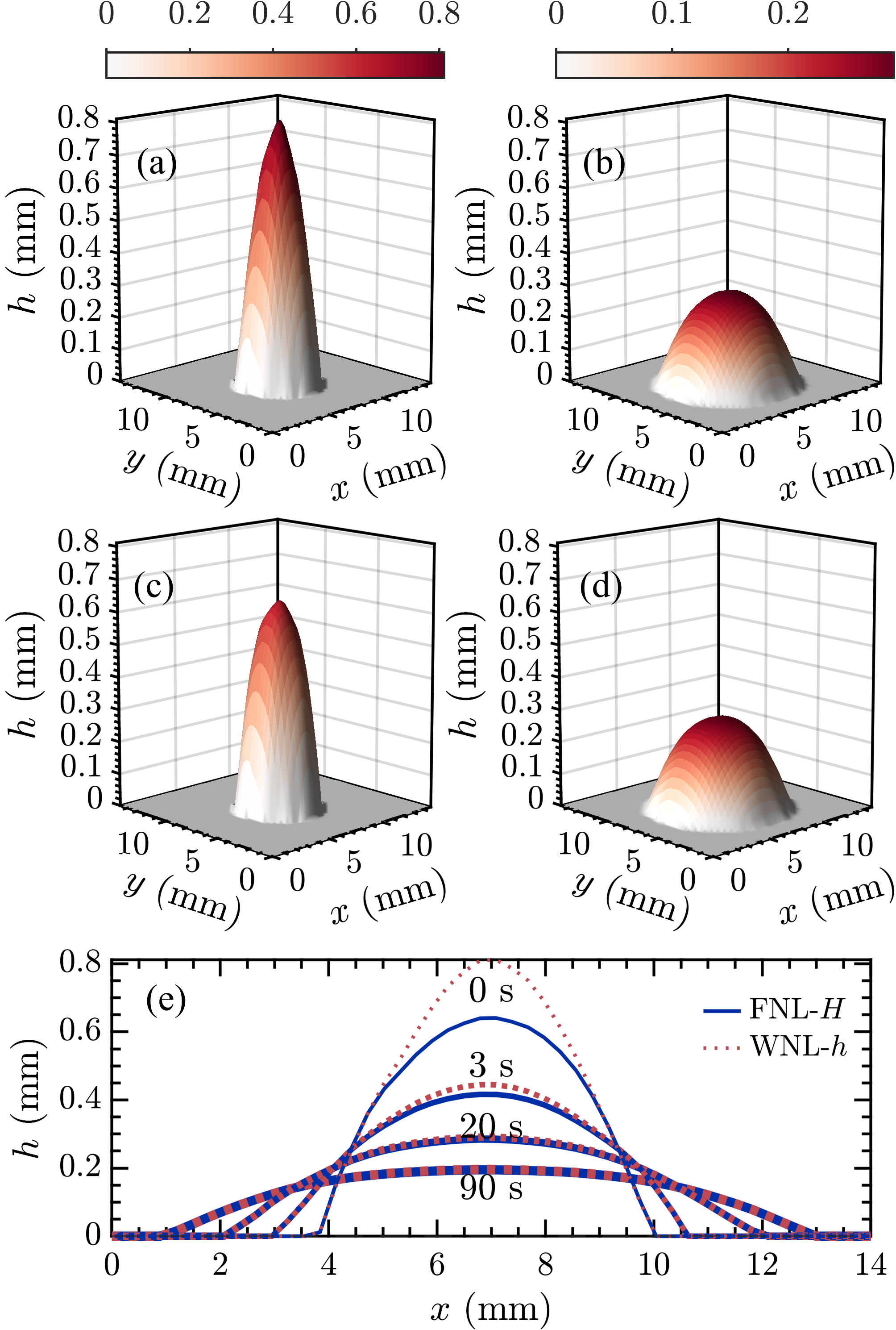}
    \caption{Reconstructed surface topography for a spreading silicone oil drop. Comparison of the 3-D surface reconstructions obtained via the (left-column) WNL-$h$ and (right-column) FNL-$H$ methods at (a, c) the early state ($t = 0$~s) and (b, d) after significant spreading ($t = 20$~s). The respective color scales at each time are provided above the panels. (e) Longitudinal surface profiles $h(x)$ extracted across the drop apex, comparing the reconstruction performance of the two methods at $t = 0$, $3$, $20$, and $90$~s}
    \label{fig:drop}
\end{figure}

The reconstructed drop topographies given by the WNL-$h$ and FNL-$H$ algorithms are shown in Fig.~\ref{fig:drop}. Panels (a-d)  reveal a substantial difference  at the early stage ($t = 0$~s, defined by the first captured frame), a difference that diminishes as the drop spreads ($t = 20$~s). This behavior is more explicitly captured in the apex profiles, $h(x)$, as depicted in Fig.~\ref{fig:drop}(e), where we see that the WNL-$h$ method significantly overestimates the peak height during the initial phase. This deviation occurs because the ratio $h/H_{\mathrm{eff}}$ is at its maximum, in which case we have a theoretical breakdown of the small-amplitude approximation that is fundamental in the WNL-$h$ method. As the drop flattens and the height-to-depth ratio recedes, the WNL-$h$ result converges toward the more accurate FNL-$H$ solution

\begin{figure}[htbp]
    \centering
    \includegraphics[width=\linewidth]{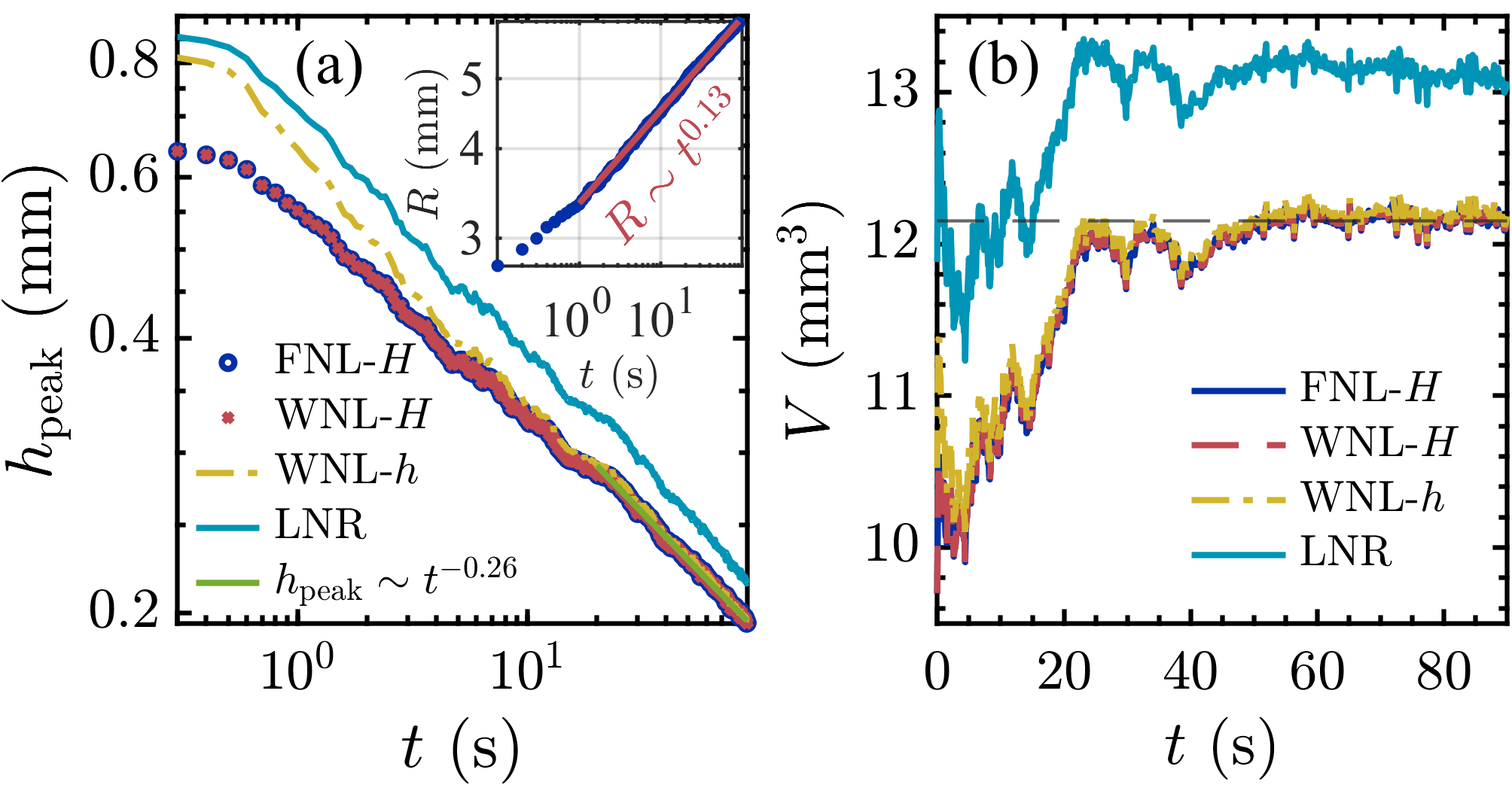}
    \caption{Temporal evolution of the spreading process measured via the four methods. (a) Decay of the maximum drop elevation (apex height) $h_{\mathrm{peak}}$. Inset: Power-law scaling of the spreading radius $R \sim t^{0.13}$. (b) Estimation of the reconstructed total volume $V$; the dashed line indicates the stable average volume of $12.15$~mm$^3$ for $t > 50$~s}
    \label{fig:drop_evolution}
\end{figure}

Figure~\ref{fig:drop_evolution}(a) depicts the temporal decay of the maximum elevation $h_{\mathrm{peak}}$ as reconstructed by the four different algorithms. The original, linear reconstruction (LNR) strongly overestimates the drop height for all times. This is due to the fact that the ratio $h_{\mathrm{peak}}/H_{\mathrm{eff}}$ remains quite high, above 0.2. As spreading progresses and the ratio $h/H_{\mathrm{eff}}$ decreases, the WNL-$h$ result gradually converges toward the high-accurate FNL-$H$ and WNL-$H$ solutions. We note that the WNL-$H$ and FNL-$H$ methods yield nearly identical results in this case. Due to the high wettability of silicone oil on glass, the surface slopes remain small throughout the process. Even at the initial stage ($t=0$), the maximum slope at the contact line is approximately $0.34$ (maximum angle 19$^\circ$), ensuring the system stays within the small-slope approximation necessary for the WNL-$H$ framework.

As shown in the inset of Fig.~\ref{fig:drop_evolution}(a), the radius of the spreading drop footprint, extracted from the non-zero displacement domain, follows a power law $R \sim t^{0.13}$. This is in excellent agreement with the classical scaling for spreading droplets in the viscous-gravity regime, typically $t^{1/8} \approx t^{0.125}$ \citep{cazabat1986dynamics} or $t^{1/7} \approx t^{0.143}$ \citep{ehrhard1993experiments}. Additionally in Fig.~\ref{fig:drop_evolution}, we observe a power-law height decay of $h_{\mathrm{peak}} \sim t^{-0.26}$ during the late-time spreading period  ($t > 20\ \mathrm{s}$). These two power laws are consistent with volume conservation: for a droplet height of the order of the capillary length (approximately $1.5$~mm for silicone oil), the interface is well-approximated as a spherical cap of volume $V = \pi h_{\mathrm{peak}}(3R^2 + h_{\mathrm{peak}}^2)/6$ where $R$ denotes the spreading radius. In the late-time flattened regime ($t > 20$~s), when $h_{\mathrm{peak}} \ll R$, one has the simplified relation $V \approx \pi R^2 h_{\mathrm{peak}}/2$, yielding $h_{\mathrm{peak}} \sim R^{-2} \sim t^{-0.26}$, in remarkable agreement with the $H$-based reconstruction results, further validating the accuracy of the iterative frameworks.

As a consistency check of the reconstruction, we can verify to what extent the measured volume of the droplet is conserved.  The result is shown in Fig.~\ref{fig:drop_evolution}(b). While the LNR method systematically overestimates the volume by approximately $1$~mm$^3$, the WNL-$h$ and $H$-based methods demonstrate superior robustness, yielding a stable estimate of $12.15$~mm$^3$ in the late-time spreading phase. A transient volume underestimation is observed with all algorithms  for $t < 25$~s and this has two reasons. First, spatial discretization errors are significant given the initially small droplet radius $R \approx 2.74\ \mathrm{mm}$. With a spatial resolution of $\Delta L = 0.295\ \mathrm{mm}$, the droplet area covers approximately 270 grid points, leading to sampling errors in the numerical calculation of the volume. Second, and more critically, the DIC algorithm smooths the sharp displacement discontinuity at the contact line across the interrogation window. Mathematically, since the effective plate thickness $H_{\mathrm{eff}}$ is finite, the displacement field, given by $\delta\mathbf{r} \propto (H_{\mathrm{eff}} + h)\nabla h$, undergoes an abrupt jump from zero on the dry substrate to a finite value at the droplet edge. This transition cannot be accurately captured by the Eulerian-based DIC used here, as the finite-size interrogation windows act as a spatial filter that ``rounds off" the discontinuity. This measurement bias naturally diminishes as the droplet spreads and the contact angle decreases (accuracy near the contact line could be further improved by employing Lagrangian-based techniques, such as particle tracking velocimetry).

\subsection{Faraday waves in a brimful square cylinder}

In a third experiment, we measure the surface topography of Faraday waves, which are parametric standing waves that appear on the free surface of a fluid layer subjected to vertical vibration beyond a critical acceleration threshold \citep{miles1990parametrically}. Due to a subharmonic instability, these waves typically oscillate at half the driving frequency of the shaker. The resulting surface patterns, ranging from simple stripes to complex lattices, provide a rigorous test case for our reconstruction frameworks due to their well-defined periodicity and significant local slopes.

The experimental cell consisted of a brimful square cylinder with inner dimensions of $60 \times 60 \times 2$~mm$^3$, fully within the limited field of view of the telecentric lens. The assembly was constructed from two stacked acrylic plates: the lower plate served as the solid base ($H_b = 2$~mm), while the upper plate, featuring a $60 \times 60$~mm$^2$ square cutout, defined the liquid reservoir ($H_l = 2$~mm). Using water as the working fluid, the refractive index ratios were $\eta = 1/1.33$ (air--water) and $\eta_b = 1.33/1.49$ (water--acrylic), yielding an effective depth $H_{\mathrm{eff}} \approx 3.79$~mm. The system was subjected to vertical sinusoidal forcing at a frequency of $f = 12$~Hz with a peak-to-peak amplitude of 4~mm. Images were captured at a frame rate of 29~fps. Given the steady-state periodicity of the Faraday waves, the frames were phase-sorted to reconstruct the full oscillation cycle. Notably, the use of telecentric lens ensured that the vertical vibration did not introduce scaling (zoom) effects in the dot pattern, thereby eliminating potential vibration-induced errors in the displacement measurements. The reference height was enforced on a point of the square boundary and locally imposes the pinned boundary condition ($h = 0$ and $H=H_{\mathrm{eff}}$).  

\begin{figure}[tbp]
    \centering
    \includegraphics[width=\linewidth]{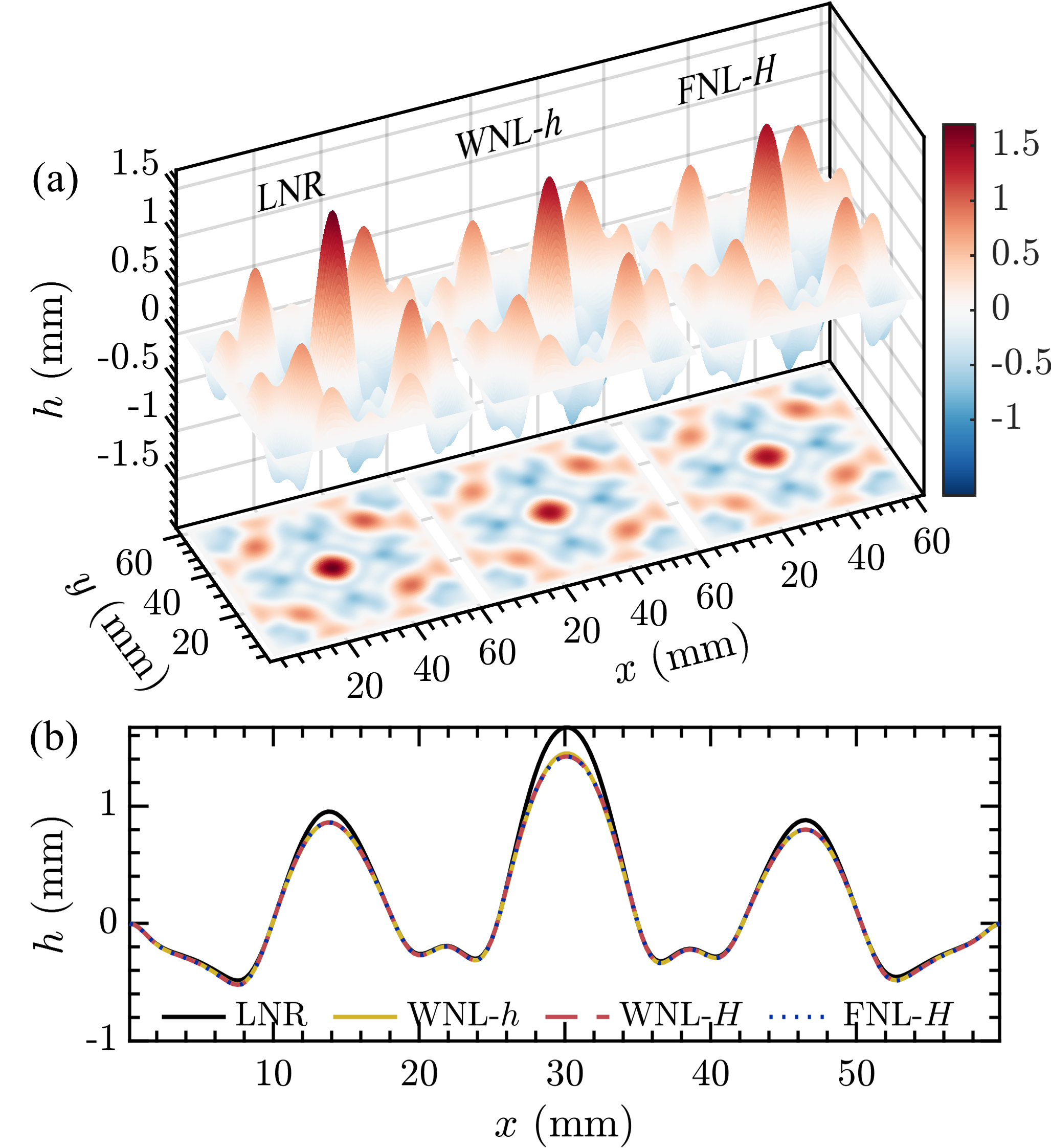}
    \caption{Reconstruction of the Faraday wave topography in a brimful square cylinder. (a) Three-dimensional surface reconstructions (top) and corresponding two-dimensional elevation maps $h(x,y)$ (bottom) for the peak frame of a wave cycle, comparing the LNR, WNL-$h$, and FNL-$H$ methods (left to right). (b) Comparison of the surface profiles extracted along the diagonal of the cylinder, highlighting the performance of the four reconstruction frameworks}
    \label{fig:Faraday_Wave}
\end{figure} 

A representative reconstruction at the phase of maximum displacement (peak frame) is presented in Fig.~\ref{fig:Faraday_Wave}(a). The 3-D surface maps and 2-D elevation fields $h(x,y)$ illustrate a clear progression in reconstruction accuracy across the different methods. Notably, the LNR method significantly overestimates the peak elevation at the center of the container, whereas the WNL-$h$ and FNL-$H$ methods yield very similar and hence, physically more reliable results. This quantitative refinement is more evident in  the surface profiles extracted along the diagonal of the cylinder shown in Fig.~\ref{fig:Faraday_Wave}(b). At the wave extrema, the local elevation-to-depth ratio reaches $|h|/H_{\mathrm{eff}} \approx 0.37$. While this value lies well beyond the linear regime ($|h|/H_{\mathrm{eff}} \ll 1$), it remains within the theoretical convergence limit ($< 0.5$) of the third-order expansion. Consequently, the WNL-$h$ and iterative $H$-based methods both demonstrate a marked reduction in reconstruction error compared to the LNR results. The largest discrepancy between the WNL-$h$ and FNL-$H$ methods emerges at the wave peaks, where the WNL-$h$ method slightly overestimates the elevation by approximately $0.03$~mm. Furthermore, the WNL-$H$ and FNL-$H$ methods yield nearly indistinguishable profiles. This is because the surface slopes $\|\nabla h\|$ remain sufficiently small such that the small-slope approximation inherent to WNL-$H$ remains valid.

Critically, the improvement of these nonlinear frameworks over the classic LNR method is most evident in the surface height quantification. Nevertheless, for this experiment, the surface reconstruction achieved by the standard LNR method still  captures correctly the spatial pattern of the wave. This is because the measured displacement $\delta\mathbf{r}$ inherently contains the signature of the wave field across all orders. Since $\delta\mathbf{r}$ is not perturbatively expanded into different orders of $\varepsilon$ and mapped directly onto the gradient field to calculate $h^{(1)}$ (Eq.~\ref{eq:h1}), the linear reconstruction naturally preserves the full spatial spectral information of the surface. Consequently, higher-order corrections in the perturbative expansion act primarily as localized amplitude redistributions, flattening peaks and deepening troughs, without fundamentally altering the underlying spatial pattern. While the LNR method effectively captures the modal structure, the WNL-$h$ and $H$-based frameworks are essential for accurate topographical characterization necessary for investigating nonlinear wave dynamics.

\section{Conclusion\label{sec:conclusion}}
In this study, we have developed and validated a hierarchy of nonlinear free-surface reconstruction algorithms, {summarized in Fig.~\ref{fig:workflow},} that systematically overcome the accuracy limitations of traditional linear FS-SS models \citep{moisy2009synthetic}. By employing a telecentric optical system to enforce orthographic projection, we effectively eliminate paraxial distortions and reduce environmental optical noise. This simplifies the light-ray displacement model, enabling the development of two iterative frameworks: $h$-based perturbative methods (LNR and WNL-$h$) and $H$-based iterative methods (WNL-$H$ and FNL-$H$). These algorithms progressively incorporate higher-order refraction terms and depth-dependent corrections, extending the validity of FS-SS into regimes characterized by large amplitude-to-depth ratios ($|h|/H_{\mathrm{eff}}$) and steep gradients ($\|\boldsymbol{\nabla} h\|$). {Convergence is typically achieved in less than 5 iterations.} 

A central finding is the significant accuracy improvement of our nonlinear methods over the classic LNR approach, as verified by both numerical and experimental tests. While $h$-based methods are computationally efficient for $|h|/H_{\mathrm{eff}} < 0.5$, they lose accuracy as the local elevation becomes a significant fraction of the effective surface-pattern distance. In contrast, $H$-based frameworks maintain high topographical fidelity by utilizing the instantaneous local depth ($H = H_{\mathrm{eff}} + h$) within the refraction model, ensuring precision within a wider range of $|h|/H_{\mathrm{eff}} \lesssim 1$.  {We also demonstrated that proper displacement interpolation is essential when using these nonlinear methods.}

Just like the original linear FS-SS method, the proposed nonlinear reconstruction algorithms are constrained by the physical limits associated with caustic formation, the accuracy of the underlying DIC data, and the requirement of at least one point of known height within the field of view. Apart from that, our nonlinear algorithms represent a significant advancement in high-resolution interfacial metrology. To facilitate broader adoption and ensure reproducibility, we provide MATLAB and Python libraries implementing these iterative reconstruction frameworks at \url{https://github.com/zhang-shimin/NLSS}.

\backmatter
\bmhead{Author Contributions}
S.Z. and W.H. conceived the study and performed formal analysis. S.Z. and F. M. mounted and conducted experimental investigations. S.Z., F.M. and W.H. developed the methodology. S.Z. and W.H. wrote the main manuscript text. W.H., F.M., and Z.L. supervised the research. All authors reviewed and approved the final manuscript.

\bmhead{Acknowledgements}
We thank H. Li, M. Avila, and D. Xu for generously sharing the code for their weakly nonlinear methods. We thank J.~Amarni, S.~Mergui, and R.~Pidoux for experimental help, and B.~Dhote, E.~Le Ster and M. Rabaud for fruitful discussions. This work was supported by the project “TransWaves” of the French National Research Agency (Grant No. ANR-24-CE51-3840-01) and the National Natural Science Foundation of China (Grant No. 52371287). The first author, S.~Zhang, also acknowledges the financial support provided by the China Scholarship Council (CSC).


\section*{Declarations}
\bmhead{Conflict of interest}
The authors declare that they have no conflict of interest.




\begin{thebibliography}{26}
\providecommand{\natexlab}[1]{#1}
\providecommand{\url}[1]{{#1}}
\providecommand{\urlprefix}{URL }
\providecommand{\doi}[1]{\url{https://doi.org/#1}}
\providecommand{\eprint}[2][]{\url{#2}}
 \bibcommenthead

\bibitem[{Abella and Soriano(2019)}]{abella2019detection}
Abella AP, Soriano MN (2019) Detection and visualization of water surface three-wave resonance via a synthetic schlieren method. Physica Scripta 94(3):034006

\bibitem[{Berhanu and Falcon(2013)}]{berhanu2013space}
Berhanu M, Falcon E (2013) Space-time-resolved capillary wave turbulence. Physical Review E 87(3):033003

\bibitem[{Cazabat and Stuart(1986)}]{cazabat1986dynamics}
Cazabat A, Stuart MC (1986) Dynamics of wetting: effects of surface roughness. The Journal of Physical Chemistry 90(22):5845--5849

\bibitem[{Cobelli et~al.(2009)Cobelli, Maurel, Pagneux, and Petitjeans}]{cobelli2009global}
Cobelli PJ, Maurel A, Pagneux V, et~al (2009) Global measurement of water waves by fourier transform profilometry. Exp Fluids 46(6):1037--1047

\bibitem[{Damiano et~al.(2016)Damiano, Brun, Harris, Galeano-Rios, and Bush}]{damiano2016surface}
Damiano AP, Brun PT, Harris DM, et~al (2016) Surface topography measurements of the bouncing droplet experiment. Exp Fluids 57(10):1--7

\bibitem[{Eddi et~al.(2009)Eddi, Fort, Moisy, and Couder}]{eddi2009unpredictable}
Eddi A, Fort E, Moisy F, et~al (2009) Unpredictable tunneling of a classical wave-particle association. Physical review letters 102(24):240401

\bibitem[{Ehrhard(1993)}]{ehrhard1993experiments}
Ehrhard P (1993) Experiments on isothermal and non-isothermal spreading. Journal of Fluid Mechanics 257:463--483

\bibitem[{Falcon and Mordant(2022)}]{falcon2022experiments}
Falcon E, Mordant N (2022) Experiments in surface gravity--capillary wave turbulence. Annual Review of Fluid Mechanics 54(1):1--25

\bibitem[{Gomit et~al.(2013)Gomit, Chatellier, Calluaud, and David}]{gomit2013free}
Gomit G, Chatellier L, Calluaud D, et~al (2013) Free surface measurement by stereo-refraction. Exp Fluids 54(6):1540

\bibitem[{Gomit et~al.(2022)Gomit, Chatellier, and David}]{gomit2022free}
Gomit G, Chatellier L, David L (2022) Free-surface flow measurements by non-intrusive methods: a survey. Experiments in Fluids 63(6):94

\bibitem[{Haudin et~al.(2016)Haudin, Cazaubiel, Deike, Jamin, Falcon, and Berhanu}]{haudin2016experimental}
Haudin F, Cazaubiel A, Deike L, et~al (2016) Experimental study of three-wave interactions among capillary-gravity surface waves. Physical Review E 93(4):043110

\bibitem[{Kurata et~al.(1990)Kurata, Grattan, Uchiyama, and Tanaka}]{kurata1990water}
Kurata J, Grattan K, Uchiyama H, et~al (1990) Water surface measurement in a shallow channel using the transmitted image of a grating. Rev Sci Instrum 61(2):736--739

\bibitem[{Li et~al.(2021)Li, Avila, and Xu}]{li2021single}
Li H, Avila M, Xu D (2021) A single-camera synthetic schlieren method for the measurement of free liquid surfaces. Exp Fluids 62:1--15

\bibitem[{Mathis et~al.(2026)Mathis, Cazin, Methel, Charru, Magnaudet, Moisy, and Rabaud}]{mathis2026effet}
Mathis R, Cazin S, Methel J, et~al (2026) Effect of wind turbulence on wave generation over a viscous liquid. arxiv p 2601.20396

\bibitem[{Mermelstein and Dagan(2025)}]{mermelstein2025modified}
Mermelstein H, Dagan Y (2025) Modified free-surface synthetic schlieren method to adjust measurement sensitivity in high-strain waves. Exp Fluids 66(1):4

\bibitem[{Metzmacher et~al.(2022)Metzmacher, Lagubeau, Poty, and Vandewalle}]{metzmacher2022double}
Metzmacher J, Lagubeau G, Poty M, et~al (2022) Double pattern improves the schlieren methods for measuring liquid--air interface topography. Exp Fluids 63(8):120

\bibitem[{Miles et~al.(1990)Miles, Henderson et~al.}]{miles1990parametrically}
Miles J, Henderson D, et~al (1990) Parametrically forced surface waves. Annual review of fluid mechanics 22(1):143--165

\bibitem[{Moisy et~al.(2009)Moisy, Rabaud, and Salsac}]{moisy2009synthetic}
Moisy F, Rabaud M, Salsac K (2009) A synthetic schlieren method for the measurement of the topography of a liquid interface. Exp Fluids 46(6):1021--1036

\bibitem[{Moisy et~al.(2012)Moisy, Michon, Rabaud, and Sultan}]{moisy2012cross}
Moisy F, Michon GJ, Rabaud M, et~al (2012) Cross-waves induced by the vertical oscillation of a fully immersed vertical plate. Phys Fluids 24(2):022110

\bibitem[{Raffel(2015)}]{raffel2015background}
Raffel M (2015) Background-oriented schlieren (bos) techniques. Exp Fluids 56:1--17

\bibitem[{Stamhuis and Thielicke(2014)}]{stamhuis2014pivlab}
Stamhuis E, Thielicke W (2014) Pivlab--towards user-friendly, affordable and accurate digital particle image velocimetry in matlab. Journal of open research software 2(1):30

\bibitem[{Vinnichenko et~al.(2025)Vinnichenko, Pushtaev, Rudenko, Plaksina, and Uvarov}]{vinnichenko2025background}
Vinnichenko NA, Pushtaev AV, Rudenko YK, et~al (2025) Background-oriented schlieren with image processing based on phase-shifting profilometry. Exp Fluids 66(3):47

\bibitem[{Wildeman(2018)}]{wildeman2018real}
Wildeman S (2018) Real-time quantitative schlieren imaging by fast fourier demodulation of a checkered backdrop. Exp Fluids 59:97

\bibitem[{Zhang et~al.(2023{\natexlab{a}})Zhang, Hector, Rabaud, and Moisy}]{zhang2023wind}
Zhang J, Hector A, Rabaud M, et~al (2023{\natexlab{a}}) Wind-wave growth over a viscous liquid. Phys Rev Fluids 8(10):104801

\bibitem[{Zhang et~al.(2023{\natexlab{b}})Zhang, Borthwick, and Lin}]{zhang2023pattern}
Zhang S, Borthwick AGL, Lin Z (2023{\natexlab{b}}) Pattern evolution and modal decomposition of faraday waves in a brimful cylinder. J Fluid Mech 974:A56

\bibitem[{Zhang et~al.(2025)Zhang, Yang, Borthwick, and Lin}]{zhang2025free}
Zhang S, Yang H, Borthwick AG, et~al (2025) Free-surface topography measurements of fluid layers over a smoothly varying bed. Exp Fluids 66(11):199

\end{thebibliography}

\end{document}